\documentclass[sigconf, nonacm]{acmart}
\usepackage{booktabs}
\usepackage{pifont}
\usepackage{tikz}
\usepackage{amsmath}

\usepackage{multirow, makecell}

\usepackage{xurl}

\usepackage{graphics}
\usepackage{array}

\newcolumntype{P}[1]{>{\centering\arraybackslash}p{#1}}
\newcolumntype{M}[1]{>{\centering\arraybackslash}m{#1}}

\usepackage{cancel}

\author{Nils Weissgerber}
\affiliation{%
   \institution{Fraunhofer FKIE}
   \city{}
   \country{}}
\author{Thorsten Jenke}
\affiliation{%
   \institution{Fraunhofer FKIE}
   \city{}
   \country{}}
\author{Elmar Padilla }
\affiliation{%
   \institution{Fraunhofer FKIE}
   \city{}
   \country{}}
\author{Lilli Bruckschen}
\affiliation{%
   \institution{Fraunhofer FKIE}
   \city{}
   \country{}}

\setcopyright{none} % No copyright notice required for submissions
\begin{document}
\title{Open \textit{SESAME}: Fighting Botnets with Seed Reconstructions of Domain Generation Algorithms}

\begin{abstract}
  An important aspect of many botnets is their capability to generate pseudorandom domain names using Domain Generation Algorithms (DGAs). A cyber criminal can register such domains to establish periodically changing rendezvous points with the bots. DGAs make use of seeds to generate sets of domains. Seeds can easily be changed in order to generate entirely new groups of domains while using the same underlying algorithm. While this requires very little manual effort for an adversary, security specialists typically have to manually reverse engineer new malware strains to reconstruct the seeds. Only when the seed and DGA are known, past and future domains can be generated, efficiently attributed, blocked, sinkholed or used for a take-down.
Common counters in the literature consist of databases or Machine Learning (ML) based detectors to keep track of past and future domains of known DGAs and to identify DGA-generated domain names, respectively.
However, database based approaches can not detect domains generated by new DGAs, and ML approaches can not generate future domain names. 

In this paper, we introduce \textit{SESAME}, a system that combines the two above-mentioned approaches and contains a module for automatic Seed Reconstruction, which is, to our knowledge, the first of its kind. It is used to automatically classify domain names, rate their novelty, and determine the seeds of the underlying DGAs. \textit{SESAME} consists of multiple DGA-specific \textit{Seed Reconstructors} and is designed to work purely based on domain names, as they are easily obtainable from observing the network traffic. We evaluated our approach on 20.8 gigabytes of DNS-lookups. Thereby, we identified 17 DGAs, of which 4 were entirely new to us.
\end{abstract}

% TODO: replace this section with code generated by the tool at https://dl.acm.org/ccs.cfm
% \begin{CCSXML}
% <ccs2012>
%   <concept>
%       <concept_id>10010147.10010257.10010293.10010294</concept_id>
%       <concept_desc>Computing methodologies~Neural networks</concept_desc>
%       <concept_significance>500</concept_significance>
%       </concept>
%   <concept>
%       <concept_id>10002978.10002997.10002998</concept_id>
%       <concept_desc>Security and privacy~Malware and its mitigation</concept_desc>
%       <concept_significance>500</concept_significance>
%       </concept>
%  </ccs2012>
% \end{CCSXML}

% \ccsdesc[500]{Computing methodologies~Neural networks}
% \ccsdesc[500]{Security and privacy~Malware and its mitigation}
% -- end of section to replace with generated code


\maketitle

%-------------------------------------------------------------------------------
\section{Introduction}\label{sec:intro}
%-------------------------------------------------------------------------------   
          
\begin{figure}[t!]\centering
\includegraphics[width=0.45\textwidth]{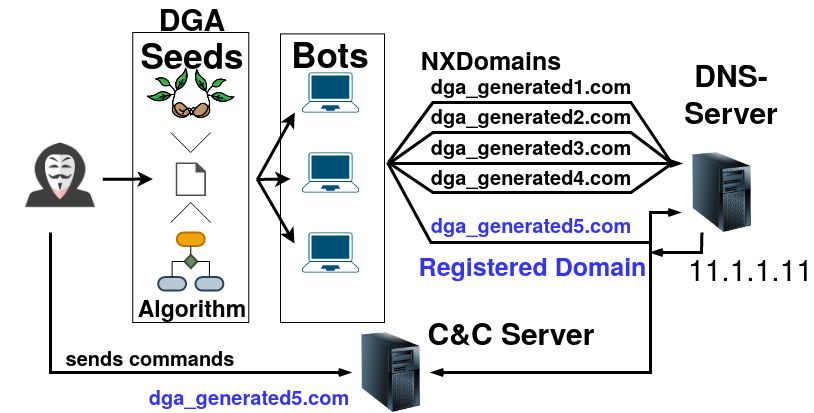}
\caption{Schematic representation of the threat model. An adversary creates malware containing a DGA. Once a system is compromised it is turned into a bot, trying to resolve domain names generated by the DGA depending on the specific seed of the DGA. The attacker knows the seed and algorithm and can register a single domain name to communicate with the bots through a C2 server.}
\label{fig:dga_illus}
\end{figure}

% General Introduction to DGAs
Domain Generation Algorithms (DGAs) were discovered in 2006 \cite{2015Botconf} and have been used by adversaries ever since as a means to increase the availability of command and  control (C\&C) servers for bots. 
DGAs can be countered by reverse engineering the malware. Once the algorithm as well as the seeds are known, future domains can be generated in advance. This knowledge can be leveraged to attribute domains to malware, employ sinkholes, block domains or help perform take-downs \cite{2020LepochatAvalanche}. A threat model is schematically shown in Figure \ref{fig:dga_illus}.

% More details about DGAs, versions, campaigns + effort asymmetry
At the time of writing, there are 99 different families of DGAs known to the public \cite{2016Dgarchive}. For many of these DGAs, multiple versions and campaigns exist. New versions use slightly modified source code to generate different sets of domains. 
A new campaign describes a new set of seeds for the same algorithm and can be created effortlessly by an adversary simply by changing one parameter. It is possible that there is more than one campaign of a DGA active at the same time. Some DGA families are known to comprise over 100 different campaigns (e.g. \textit{Locky})\cite{2016Dgarchive}. 
The majority of DGAs have to be manually reverse engineered to extract the seed in order to interfere with the campaign. Depending on the particular malware sample, this may prove to be a difficult and/or a very time consuming task as obfuscation can be used to harden the code against reverse engineering (e.g. \textit{Nymaim} \cite{webCertPlNymaim})\cite{2020REHandbook}\cite{2007LimitsOfRe}. This asymmetry in effort between the adversary and security specialists is unfavorable. Therefore, an effortless and automatic method for domain attribution, detection, and seed reconstruction is required.

% DGArchive and Popularity of DGAs
In 2015, Plohmann et al. \cite{2016Comprehensive} investigated 43 different DGAs with more than 253 seeds and used the results to introduce a taxonomy for different types of DGAs. The authors also published a database called DGArchive \cite{2016Dgarchive}, where information about DGAs and their seeds is accumulated. The information is used to generate domains for all DGA versions and campaigns in their database ahead of their validity periods, i.e. the time span criminals are using these domains for C\&C communication. Since the aforementioned publication, the number of DGAs rose to 99 (+130\%) with over 840 seeds (+232\%), showing the ongoing importance of DGAs.

% Two approaches to DGA counters, one dgarchive
There are two main approaches for countering DGAs. On the one hand, the information from databases such as DGArchive can be used. However, this only works for known algorithms and seeds. Furthermore, a malware sample and reverse engineering is required to extract those. 

% Machine Learning efforts for (unknown) AGD Detection
On the other hand, known and unknown Algorithmically Generated Domains (AGDs) can be automatically detected through machine learning. For this, a multitude of detection methods have been suggested (e.g.\cite{2010DetectAgd}, \cite{2013Tracking}, \cite{2016Woodbridge}, \cite{2019Wordbased}). Until 2016, many publications made use of handcrafted features for AGD detection (e.g. \cite{2013Tracking},\cite{2010DetectAgd},\cite{2012Throaway}, \cite{2014Phoenix}). More recently, Machine Learning (ML) is primarily applied for this task (e.g. \cite{2016Woodbridge}, \cite{2017LSTMimbalance}, \cite{2017SupLearn}, \cite{2020LepochatAvalanche}). ML models appear to achieve great success in correctly differentiating between AGDs and Humanly Generated Domains (HGDs). This holds, even for domains which are unknown to databases like DGArchive or even the particular ML model. 

However, this technique reaches its limits quickly as it only allows filtering/blocking domain lookups but does not help in predicting future domains. The asymmetry of effort between the adversary, to change a seed, and the security specialist, to reconstruct it, remains. Therefore, seed reconstruction or prediction is necessary to substantially lower the manual amount of effort required for the security specialists.

% Our work
In this paper, we tackle the shortcomings of the two approaches and introduce our \textit{SESAME} system (\textit{SE}ed reconstruction from \textit{SA}mples by \textit{M}ulticlass \textit{E}valuation). It consists of two modules: the \textit{Batch Classifier} and \textit{Seed-Reconstructor}. The former discovers new DGAs, versions and campaigns in DNS data given as input by classifying domain names and calculating a suspicion score, which quantifies the level of similarity to known domains. This way, no malware sample is required. Subsequently, the \textit{Seed-Reconstructor} reconstructs seeds from AGDs automatically without the need of additional manual reverse engineering or human intervention, and thus equalizes the effort-asymmetry between criminals and security specialists. To show the capabilities of \textit{SESAME}, we evaluated it on over 20 GB of real world DNS data. These DNS-lookups are generated by sandbox runs of real world malware and are generously provided by the ShadowServer Foundation\cite{shadowserver}.\\
In summary, we make the following contributions:

\begin{itemize}
    \iffalse
    \item We present an approach named \textit{Batch Classifier}, for classifying and attributing domain names from DNS data as well as assigning scores that indicate the degree of novelty of these domain names.
    %This is done by a combination of a trained machine learning model and existing regular expressions for known DGAs and seeds.
    \fi
    \item We make use of a state-of-the-art Neural Network to classify and attribute domain names from DNS data. Furthermore, we build upon these classifications by calculating scores that indicate the degree of novelty of these domain names in our \textit{Batch Classifier} module.
    \item We present a novel approach to automatically determine seeds based on DGA names and domain names via \textit{Seed-Reconstructors}. %These take the result of the \textit{Batch Classifier} as input and use different reconstruction techniques to determine the seed for given AGDs.
    \item We implemented and evaluated \textit{SESAME}, a system combining the \textit{Batch Classifier} and the \textit{Seed-Reconstructors} to automatically classify DNS data and reconstruct seeds from AGDs. We evaluated \textit{SESAME} with 20 GB real world DNS data from sandbox runs of malware executed between January 2018 and December 2020.
\end{itemize}

%-------------------------------------------------------------------------------
\section{Related Work} \label{sect:related_work}
%-------------------------------------------------------------------------------
One of the first AGD detection methods was published by Yadav et al. \cite{2010DetectAgd}. This method makes use of the entropy in uni- and bigrams along with the character distributions in domain names to differentiate between AGDs and HGDs. Uni- and bigrams are a sequence of one or two consecutive characters respectively.  

Antonakakis et al. \cite{2012Throaway} created a system called \textit{Pleiades} which combines clustering with a Hidden Markov Model. It is used to find clusters of similar domains in NXDomain responses based on a set of handcrafted features. In a second step, the clusters of domains are attributed to known DGAs.

Multiple papers have been published since then, focusing on the detection of AGDs. Many of these perform the detection without the use of machine learning (e.g. \cite{2014LenDistrib, 2014Phoenix, 2014Exposure, 2016SemanticAna}). 

Recent papers shifted towards the usage of machine learning algorithms as the results have proven to improve detection performance (\cite{2017DetectBroadLen}\cite{2017SupLearn}\cite{2018_embed}). For example, Sch\"uppen et al. \cite{2018Fanci} proposed the system \textit{FANCI} which is based on Random Forests (RFs) and Support Vector Machines (SVMs) and uses 21 handcrafted features extracted from the domain names of NXDomain DNS traffic. By employing an additional intelligence module, the authors were also able to find entirely new DGAs in their gathered data.

Woodbridge et al. \cite{2016Woodbridge} introduce a Recurrent Neural Network (RNN). In particular, they use a Long Short Term Memory (LSTM) network architecture \cite{1997LstmOrigin} for AGD detection. LSTMs do not require handcrafted features, instead they work on the domain strings directly. 

Recent works from Drichel et al. \cite{2020AachenLstm} propose classifiers based on residual neural networks (ResNets) for DGA detection. They show a slightly increased performance of the ResNets over LSTM models and models based on Convolutional Neural Networks (CNN) while decreasing the training time. Most papers focused on the binary problem of differentiating between HGDs and AGDs. However, the authors of this paper compare both binary class (HGDs or AGDs) and multi-class (Specific DGA family) models.

Drichel et al. \cite{2020AachenImbalance} investigated the well known problem of class imbalance (\cite{2000class}\cite{2006NNImbalance}\cite{2009ClassImbalance}\cite{2015classImbalanceReview}) in supervised machine learning, both for the binary and multi-class case. For this, they compared classification performance of SVMs, RFs, Recurrent Neural Nets (RNNs), CNNs, and ResNet-based classifiers. They concluded that including samples of underrepresented classes can lead to the correct classification of several of these, while only barely decreasing the overall performance. For this, the classes have to be made cost-sensitive which is done by applying class weights as suggested by Tran et al. \cite{2017LSTMimbalance}. 

Supervised machine learning algorithms require sets of data to learn from. These data sets must be well-chosen (i.e. representative for the test data) to achieve good performances. In the binary classification problem of differentiating between AGDs and HGD, typically one set of data contains only AGDs while the data set containing the HGDs is commonly taken from public sources such as the \textit{Alexa}-1-Million data set\cite{webAlexa}. According to Le Pochat et al. \cite{2019_tranco_alexa}, 133 top-tier studies published between 2014 and 2018 based their research on these public data sets. The authors continue to show how easily these data sets can be manipulated and in turn released their own service called `Tranco'\cite{webTranco}, providing a more reliable and resilient set of data. They create their data set by combining the three public data sets from \textit{Alexa}\cite{webAlexa}, Cisco \textit{Umbrella}\cite{webUmbrella} and \textit{Majestic}\cite{webMajestic} into one.

Supervised machine learning algorithms also require a data set containing AGDs. One such data set is recorded and kept up to date by DGArchive \cite{2016Comprehensive}. DGArchive contains (pre-) calculated AGDs generated by re-implementations of DGAs and their known seeds found by reverse engineering malware samples. At the time of writing, DGArchive comprises more than $159 \times 10^{6}$ AGDs and 99 different DGA families.

%-------------------------------------------------------------------------------
\section{Data} \label{sect:data_sets}
%-------------------------------------------------------------------------------

Our work is based on three sets of data: \textit{AGD} data, \textit{HGD} data and \textit{DNS-logs} data. The first two of these data sets are used for the training of the machine learning classifier. The third data set contains DNS data from sandbox malware runs and is used for the evaluation. The \textit{AGD} and \textit{DNS-logs} data originate from sources with restricted access, meaning that only users validated by the owner of the data can access the data. In the following subsections, we detail on these different sets of data and their origins.

In the remainder of this work, we refer to a domain as `known' if it is generated by currently known DGAs and their currently known seeds. Accordingly, a `known campaign' describes a campaign derived from a known DGA with known seeds, hence all generated domains are known.

%-------------------------------------------------------------------------------
\subsection{Machine Learning: Training Data} \label{subsect:mldata}
%-------------------------------------------------------------------------------
An ML model's performance is heavily dependent on the data used to train it. Therefore, special care is taken while aggregating the training data set for the model. The process is explained in the following subsections. Our model needs to be able to predict the correct DGA family for a batch of domains. Hence, the model requires sample data from every single DGA family. This raw data is gathered from the sources listed below.

%-------------------------------------------------------------------------------
\subsubsection{DGA Families}
%-------------------------------------------------------------------------------
Most of the sample data for the different DGA families are provided by DGArchive \cite{2016Dgarchive}. We used a full database export from June 19, 2020 which consists of 86 different DGA families with more than $86 \times 10^{6}$ unique domains. This data includes all domains generated until the end of 2020. Another 13 DGA families, along with their \textit{Python} implementations, were acquired through publicly accessible sources such as the blog \cite{webBaderBlog} of Johannes Bader and his GitHub\cite{webBaderGit} and Netlab360 sources (Website\cite{webNetlab360Website} and GitHub\cite{webNetlab360Git}). 
Lastly, one DGA that has been found during an earlier deployment of \textit{SESAME} was added to the data set such that in the following deployments this particular DGA can be identified and correctly classified.
%Lastly, one new DGA found during our analyses was also added to the data set. Investigations proved this to be the \textit{Shylock} DGA based on Yara rules recorded in Malpedia\cite{web_malpedia}.

Hence, a total of 100 different DGA families were acquired. We removed the two DGAs \textit{randomloader} and \textit{dnsbenchmark} from the training data because these DGAs generate only 5 unique domains. Thus, 98 DGAs were used to train the ML model. However, only a small subset of domains was used for the training.
For most families approximately 5000 domains were included in the training data set. This number was chosen to ensure proper representation of all versions, campaigns, TLDs and generation times while keeping training duration of the ML model and memory usage moderate. Furthermore, this approach reduces effects from class imbalance.

There are 60 DGA families which generate more than 5000 domains. For these, a filtering had to be applied to include only 5000 domains. The filtering criteria are as follows: 
\begin{itemize}
	\item Randomly pick throughout all generated domains.
	\item For time dependent families: randomly pick throughout all dates.\footnote{For the time-span from DGA first seen until end of 2020.}
	\item For families with multiple campaigns/versions: randomly pick, but equal amounts per campaign/version. %(e.g. a family with 2 campaigns would contain 2500 domains generated from each one).
	\item For families with multiple TLDs: include representative amounts of different TLDs.
\end{itemize}
For 22 DGA families that generate less than 5000 unique domains, new domains were generated in order to reach that amount or close to it. Note that no new seeds were `created' in order to generate new domains, because this would hinder the ability to detect unknown campaigns later on. Instead, the number of generated domains for known seeds was increased. This can be done for DGAs which use Pseudo-Random-Number-Generators (PRNG)\cite{webPrng} without problems. PRNGs generate reproducible output depending on an input seed. 

This leaves 15 families for which the number of generated domains cannot be increased to reach 5000 domains. For these families, all unique domains were included.

Deviations from this number are handled automatically by utilizing a cost-sensitive model (details in Sect. \ref{sect:mlc}).

%-------------------------------------------------------------------------------
\subsubsection{`benign' Data}\label{subsub_benign_data}
%-------------------------------------------------------------------------------

One additional class (consisting of 5000 domains) is added to the 98 DGA families. This class represents the HGDs - the domains which are not generated by DGAs. This set of domains is often referred to as `benign' and comprises humanly generated domains like `youtube.com'.

We decided to use the Tranco list\cite{2019_tranco_alexa} to gather this data set, as it is more resilient against attacks than the often used Alexa data set. In particular, we use the list created on 20 July 2020\cite{webTrancoDate}. However, even in this list we have spotted multiple malicious domains which we identified as the \textit{Phorpiex} malware (i.e. domain `tefiefijiejdijef.ws'). In an attempt to reduce the contained malicious domains, we limit our usage of this data set to the first 500,000 domains.

We tested several options to pick 5000 domains from this data set in order to maximize the model's accuracy. The first option simply included the first 5000 domains. The second option randomly picks 5000 domains throughout the data set. The last option picks every k-th domain until 5000 domains are reached for $k \in \{5,10,20,50\}$.

The first option led to the best results. This was to be expected as these top 5000 domains are the most difficult to manipulate, yet they already show enough variety for the classifier to correctly classify these non-DGA-generated domains.

%-------------------------------------------------------------------------------
\subsection{Sandbox Logs} \label{subsect:sandbox}
%-------------------------------------------------------------------------------
The trained model is used to find unknown campaigns of known DGAs within Sandbox DNS-logs, generously provided to us by the \textit{Shadowserver Foundation}\cite{shadowserver} on a daily basis. These logs are generated from executing malware samples within a secluded system and recording all following DNS-lookups. The following data is recorded: \textit{timestamp}, \textit{md5hash}, \textit{domain name}, \textit{DNS-type}, \textit{DNS-response}.

The following different DNS-types are possible:`A' (IPv4 Address),  `AAAA' (IPv6 Address), `MX' (Mail Exchanger Record), `PTR' (Reverse DNS Lookup), `TXT' (Text record). The \textit{md5hash} is the md5sum of the executed malware sample.

These are all the data fields used in order to identify unknown campaigns/versions of known DGA families. 

The feeds contain strongly varying amounts of domains every day ranging from a few $10^{5}$ up to more than $10^{6}$. At the same time, the number of domains recognised by DGArchive, the number of NXDomain responses and the number domains of each type varies similarly. However, typically the majority of the domains return an NXDomain response and are of type `A'. A final summary of all data sets and what they are used for is given in Table \ref{table:dataset_usage}.

Before discussing the \textit{SESAME} system, we want to point the reader to a detailed analysis of the Sandbox logs performed in Appendix \ref{apdx:log_details}.

\begin{table}[ht]
\centering
\begin{tabular}{|c|p{2.1cm}|p{2.9cm}|}
    \hline
    \textbf{dataset} & \textbf{used domains} & \textbf{usage} \\
    \hline
    DGArchive & \multirow{2}{*}{ Known AGD } & \multirow{3}{*}{ML Model Training}  \\
    \cline{1-1} Public Sources  &   &   \\
    \cline{1-2} Tranco List   &  Known HGD &   \\
    \hline
    Sandbox logs & Network traffic domains & ML Model Application \\
    \hline
\end{tabular}%
\caption{Summary of the used data sets. The sources for every data set are given along with the usage of their contained domain names.}
\label{table:dataset_usage}
\end{table}

%-------------------------------------------------------------------------------
\section{\textit{SESAME}} \label{sect:analysis}
\begin{figure}[t!]\centering
\includegraphics[width=0.9\columnwidth]{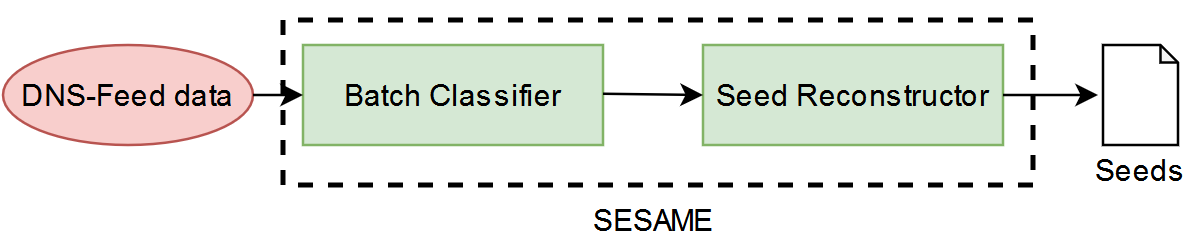}
\caption{High-level overview of the program flow of \textit{SESAME}.}
\label{fig:program_flow_SESAME}
\end{figure}

%-------------------------------------------------------------------------------
With the data mentioned in Section \ref{subsect:mldata}, a supervised ML model is trained, which is described in Section \ref{sect:mlc}. This model is used in the \textit{Batch Classifier} (Section \ref{sect:batch_classifier}), the first module of \textit{SESAME}. It analyses the Sandbox logs and identifies `suspicious' samples i.e. possible candidates for new DGAs or new DGA versions or campaigns. For applicable candidates, the second module of \textit{SESAME} is used: the \textit{Seed Reconstructor} (Section \ref{sect:seed_extr}). It leverages the gathered information about the `suspicious' samples to reconstruct the seed(s) automatically. A very high-level overview about \textit{SESAME}'s program flow is shown in Figure \ref{fig:program_flow_SESAME}. All of the modules are described in detail in the following subsections. Finally, a list of all the technical challenges and how they were solved, is given in Table \ref{table:tech_chals} in Appendix \ref{apdx: challenges}.

%-------------------------------------------------------------------------------
\subsection{Training the ML-Classifier} \label{sect:mlc}
%------------------------------------------------------------------------------- 
\textit{SESAME} uses a multi-class model to predict the family responsible for generating a domain name. This is necessary in order to identify which Seed Reconstructor to use in the second module.
%determine whether a certain group of domain names is the result of a currently known DGA and its known seeds or not.

Based on the presented related works, especially the one by Drichel et al. \cite{2020AachenLstm}, we decided to implement the ResNet model which the authors have generously provided us with. In their work, Drichel et al. showed the superiority of the ResNet to other well-performing model architectures such as an LSTM model (based on Woodbridge et al. \cite{2016Woodbridge}) and a model based on two 1-dimensional CNN layers (proposed by Yu et al. \cite{2018_yu}). In the following subsection, we discuss the model's preprocessing of data, class imbalance handling, and evaluation.

%\textcolor{gray}{\subsubsection{Machine Learning Model} \label{sect:model_decide}
%Recently, a study was published by Drichel et al. \cite{2020AachenLstm} comparing multiple state-of-the-art classifiers including an LSTM model based on the work of Woodbridge et al. \cite{2016Woodbridge} and a model based on two 1-dimensional CNN layers proposed by Yu et al. \cite{2018_yu}. They concluded that their proposed residual neural network (ResNet) outperforms all other models. Therefore, we decided to implement the ResNet which the authors have generously provided us with.}
%\textcolor{gray}{

\subsubsection{Preprocessing Training Data} \label{sect:model_preprocessing}
We converted all domain names to lowercase, mapped every character to a unique integer and kept the top-level-domains. We included a padding of domain names up to a domain length of 59 characters. This was the length of the longest domain included in the training data for the model. While actual domain names can be longer, this occurs rarely \cite{webDomainNameDistr}. Longer domain names were truncated in our approach. The advantage of this comes in shorter training times as well as less memory usage. Furthermore, we stripped all subdomains as tests have shown strong and undesired biases when classifying unknown domains with known subdomains.

\subsubsection{Class Imbalance} \label{sect:model_imbalance}
In order to counter class imbalance problems, we employed a cost-sensitive approach by applying class weights as done by \cite{2017LSTMimbalance} and \cite{2020AachenImbalance}. The class weights are calculated in the following way:
\begin{equation}
C_i = \left(\frac{N}{N_i}\right)^{\gamma} ,
\end{equation}
with the total number of samples $N$, the number of samples per class $N_i$ and a scaling factor $\gamma$. We chose $\gamma = 0.3$ as suggested by the authors of \cite{2020AachenImbalance}. The authors of \cite{2017LSTMimbalance} also found that $\gamma = 0.3$ provides good results for their RNN-based DGA classifier.

Our tests have shown that we can further improve the models performance by minimizing the number of underpopulated classes. Thus, we increased the number of generated domains for all DGAs which allowed for this (acc. to Sect. \ref{subsect:mldata}).

\subsubsection{Model Evaluation} \label{sect:model_eval}
We employed a 5-fold cross validation to test the performance of the model, when classifying data not used during training. This means we split the data set into five equally sized parts and performed five separate training procedures. In every procedure, the model is trained on four parts of the data and validated on the remaining part. During this, we made sure that each part ends up in the validation set exactly once. The individual models were discarded after each training session, but the evaluation scores were retained. The final evaluation scores were calculated by averaging the values from the five training runs.

Our final model was trained by going through two iterations, each using a different random split of the data set into a training and validation set. The training-validation split we used is 0.7:0.3. This way, the training data set contained enough data for training while lowering the chances of overfitting. At the same time, the validation set was large enough to test the versatility of the trained model. Every iteration included eight training epochs where the model reaching the highest `Area Under Curve' (AUC) is kept.
The metrics were calculated according to Appendix \ref{apdx: calcs}

%-------------------------------------------------------------------------------
\subsection{Module 1: Batch Classifier}\label{sect:batch_classifier}
%-------------------------------------------------------------------------------

%\iffalse
\begin{figure}[t!]\centering
\includegraphics[width=0.99\columnwidth]{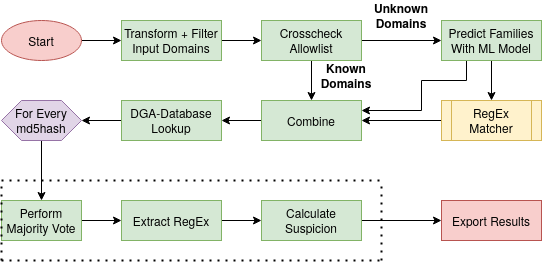}
\caption{Visualisation of the program flow of the \textit{Batch Classifier}.}
\label{fig:program_flow_batch_classifier}
\end{figure}
%\fi

The \textit{Batch Classifier} is the main computing program of our \textit{SESAME} system. It is used to filter the input for potentially interesting samples, which comprise new DGAs and new versions and campaigns of known DGAs. In this module, the DNS-feed data was transformed and classified. Additionally, a certain level of suspicion, henceforth called \textit{suspicion score}, was calculated for every analysed batch of domains. A high score indicates a group of domains following a scheme very different to the ones known to the model. This included hard-coded domains and domains from new DGA-families or new DGA versions or campaigns of already known DGAs. Therefore, groups of domains with high scores should be analysed more thoroughly. This module performs multiple steps successively in order to determine this score. The program flow is visualised in Figure \ref{fig:program_flow_batch_classifier} and detailed upon in the remainder of this Section.

\subsubsection{Preprocessing DNS-Logs data}\label{sect:batch_classifier_preprocesing}
We first transformed the domain names such that they all follow the same format of the data with which the model was trained (see Sect. \ref{sect:model_preprocessing}). Afterwards, a few filters were applied to the input data in order to remove the domain names we consider invalid. This includes removing domain names that:

\begin{itemize}
  \item Do not contain a single dot.
  \item \raggedright Contain invalid characters such as: !"§\$\%\&\_;,$<>$/()=?\{[]\}'\#+*@$\vert\sim$ .
  \item Consist of less than 4 characters.
  \item Are duplicate per malware sample.
  \item Use DNS-lookup types: `NS', `PTR' and `MX'.
\end{itemize}

Subsequently, we filtered out malware samples which contain less than 20 domains. Including a large number of domains leads to better statistical accuracy because the standard deviation gets smaller. However, at the same time, malware samples containing DGAs generating only a small number of domains may be overlooked. The number 20 is low enough to not exclude a large amount of groups of domains while offering enough domains to exploit the power of the majority vote explained in Sect. \ref{sect:batch_classifier_vote}.

Thereafter, malware samples of which less than 50\% of domain names received a NXDomain response were removed. This is because of the assumption that most of a DGAs domain names will not be registered \cite{2012Throaway}. The remaining domains are cross-checked with the first 500,000 domains from the TRANCO list serving as an allowlist. 

\subsubsection{Classification}\label{sect:batch_classifier_classification}
Domains that are not found on the allowlist are classified. During this step, every single domain will be assigned a certain family, purely based on the machine learning model. This may lead to cases where the predicted family's regular expression (RegEx) does not match the classified domain. For example, the domain `abcd89.com' may be classified as a family that can only generate domain names without digits. This may simply be a wrong classification, or the result of a RegEx change of the family. A family's RegEx can change when the malware author releases a new version of the malware in which a different set of characters or a new top-level domain is used to generate the domains.

A family's RegEx can be extracted either from the source code of a DGA or from the generated domain names. This way we created the RegExes of all the families known to the ML model. The \textit{Batch Classifier} comprises a feature we called \textit{RegEx-Matcher}. It picks the next most likely family that matches the domain's RegEx. Discarded families apply to all domains within the same group (same sample hash). This is also the reason why new DGA-families, generating domains which do not follow any previously known RegExes, will typically be classified as `benign' with the \textit{RegEx-Matcher} enabled. That is because the `benign' class has a very broad RegEx which allows for any valid domain name. The \textit{Batch Classifier} is performing two predictions, one with and one without making use of this \textit{RegEx-Matcher}.

\subsubsection{Database Lookup}\label{sect:batch_classifier_dgarchive_crosscheck}
The classified families were combined with the ones which were identified and classified through an allowlist. Every domain was marked with a tag describing how the domain was classified: `prediction' vs. `allowlist'. The domains tagged with `prediction' were not recognised by the allowlist and thus were assigned a DGA family by our ML model. The domains tagged with `allowlist' contain only `benign' domains listed on the Tranco list.

Subsequently, all the domains classified by the machine learning model were looked up in the database containing all known AGDs, for which we used the DGArchive\cite{2016Dgarchive} data. This is done because we are interested only in new versions, new campaigns and new DGAs which generate domains unknown to the database.

Note, that we checked the entire database and not just the domains generated on the execution date of the malware taken from the DNS-feed. This way, the system becomes more robust against faulty system times. However, at the same time this may lead to an inability of correctly identifying new versions of DGAs where only a time offset is introduced. This reduces the false positives while potentially increasing the false negatives. However, while analysing currently known DGAs, we have not seen occurrences where new versions introduced only a time offset, which is another reason why we opted for our approach.

\subsubsection{Majority Vote}\label{sect:batch_classifier_vote}
Another feature of the \textit{Batch Classifier} is the majority vote. Groups of domains generated by the same malware can be identified by their sample md5hash. The majority vote for each sample is performed by counting every domain's classification as single vote. The DGA family carrying the most votes in the end is assumed to be the representative family for the sample. Therefore, our classifier does not need to predict every single domain correctly in order to correctly predict the family of the entirety of the sample.

\subsubsection{RegEx Extraction}\label{sect:batch_classifier_rgx_extract}
In Sect. \ref{sect:batch_classifier_classification} we discussed how singular domain names may not match the ML model's predicted family's RegEx. In contrast, during \textit{RegEx Extraction}, the RegEx for every malware sample was automatically extracted from the entire set of domains of each sample and follows the format: \textit{[used characters]\{domain lengths\}\textbackslash.(tld1$\vert$tld2)\$} (e.g.: [a-y1-5]\{22,26\}\textbackslash.(com$\vert$net)\$). This was done in order to correctly compare it to the RegExes of the predicted families (which can only be performed after the majority vote). Assuming a correct classification, it is expected that the extracted RegEx is similar or equal to the predicted one.

Multiple RegExes can match for the same strings. E.g. the three RegExes: 
`[a-z1-9]\{10\}\textbackslash .com\$', `[az]\{10\}\textbackslash .com\$', `[\textbackslash w]\{10\}\textbackslash.com\$' all match domains of 10 characters containing the letters `a' and `z' ending on `.com'. Some RegExes can be very general while others can be very specific. Therefore, we decided to use the aforementioned format which is very general and can be applied to all DGA families.

The extracted RegEx can be affected considerably by outlier domains. These can be `benign' domains within sets of DGA domains. To make the extraction more robust against these outlier domains, we only considered domains not identified by an allowlist for the extraction. Furthermore, we removed domains whose length appear less than 5\% of the most frequently seen domain length. This is done because DGAs often generate domains with a small variation in domain length. This makes outlier domains stand out.

\subsubsection{Suspicion score}\label{sect:batch_classifier_sus_score}
The suspicion score is calculated for every malware sample. This score is used to quantify differences between already known and potentially new (to the ML model) malware samples. High values indicate a large difference between the schemes of the analysed domains and the ones known to the model.

In order to get a reliable score, it is advantageous to use as many indicators for novelty as possible. We identified four such indicators that carry information about how suspicious the analysed domains are:
\begin{enumerate}
	\item \textbf{`Benign' ratio: $A$}: ratio of number of domains classified as `benign' and total number of predictions.
	\item \textbf{Cardinality ratio: $B$}: ratio of number of unique families predicted and total number of predictions.
	\item \textbf{RegEx-change ratio: $C$}: ratio of RegEx-changes and total number of predictions \footnote{only used when the \textit{RegEx-Matcher} feature is used, average value is adjusted accordingly}. 
	\item \textbf{Mean maximum probability: $D$}: mean value of the highest probabilities from classification \footnote{The probabilities can be understood as a measure for the level of certainty of a prediction.}.
	\item \textbf{`Known' ratio: $\alpha$}: ratio of domains recognised by the DGA-database and total number of predictions.
\end{enumerate}
The higher the first three indicators are, the higher the suspicion score should be. For the last two indicators, the opposite is the case: a smaller value should lead to a higher suspicion score.

Each of the first four indicators is normalized via a root function such that they can assume values from 0 to 100. A root function was chosen so that differences for low values of the indicators have higher impact on the final score than for high values of the indicators. Since we expect that none of the first four indicators ($A$ to $D$) is stronger than the others, we assigned equal weight to each of them, simply by calculating the mean of the individual parts.

However, the last indicator ($\alpha$) is treated differently: A sample of which all domains are recognised, is irrelevant and thus, should receive a final score of 0, which might not happen if it was treated like the other indicators. On the other hand, a sample that does not appear suspicious based on the first four indicators should still receive a final score that is high enough to be considered `suspicious' if all the domains are unknown. For this, a certain base suspicion score $S_{\text{base}}$ is required. We chose $S_{\text{base}}=10$ such that, if the first four indicators lead to an average of 0 and less than 40\% of domains of the sample are recognised, a sample appears as `suspicious' (achieves a score $>$ 5.0, a value chosen based on results shown in \ref{sect:ident_tests}). Due to the base score, we divided the value by 1.1 in order to normalize the suspicion score to 100.

Since some malware queries domains unrelated to the DGA, it can happen that even a known malware does not reach a `Known' ratio of $\alpha = 1$. To account for that, we assumed ratios of up to $\alpha = 0.96$ as `Known' and assigned them a suspicion score of $S = 0$. Furthermore, some AGDs will never be recognised by looking them up in DGArchive, simply because they are not stored there. This is because the DGA is non-deterministic, meaning that every time the DGA is executed, a new set of domains is generated. So far, we know of two non-deterministic DGAs: "Reconyc" and "Shylock". In order to not automatically assign very high scores to these families, we assigned the value of $\beta = 2$ to half the suspicion score. This also means that for samples of these families to appear `suspicious', at least one other indicator is required.

Thus, we ended up with the following set of equations to calculate the suspicion score $S$:
\begin{flalign}
& S = \left( \frac{A + B + C + D}{4} + S_{\text{base}} \right) \times \frac{\gamma}{1.1} \\
& \gamma = \begin{cases}
\frac{1-\alpha}{\beta} , & \text{if } \alpha \leq 0.96\\
0              & \text{otherwise}\end{cases} ,\quad\quad S_{\text{base}} = 10\\
& \beta = \begin{cases}
1 , & \text{if deterministic DGA} \\
2              & \text{otherwise}\end{cases}&   
\end{flalign}

%-------------------------------------------------------------------------------
\subsection{Module 2: Seed-Reconstructor} \label{sect:seed_extr}
%-------------------------------------------------------------------------------
The \textit{Seed-Reconstructor} module was designed to take the resulting data of the \textit{Batch Classifier} module as input. The data contains a suspicion score and a predicted family for every analysed sample. If a sample's score surpasses a certain threshold, which can be set for every DGA family separately, the sample will automatically undergo seed-reconstruction. For most DGA families, this threshold is set to 5.00 based on tests explained in Section \ref{sect:ident_tests}. This value is small enough to include the majority of new versions/campaigns while filtering out the known ones.

Once a threshold is surpassed, the sample's domain names were used as input for the specific Seed-Reconstructor. Certain Seed-Reconstructors additionally require a date to work. In these cases, the timestamp of the analysed DNS-logs was used. 

Because DGAs are very versatile, there is no common unified method to reconstruct the seeds for every single DGA based on domain names. Hence, every DGA requires its own unique seed reconstruction technique. However, during our analysis of the DGAs, multiple groups of Seed-Reconstructor types could be established. There are four different types, all of which appear similarly often, as indicated by the number behind each group:
\begin{enumerate}
\item Permutators \& Iterators (5 Reconstructors)
\item Bruteforcers (7 Reconstructors)
\item Smart Bruteforcers (8 Reconstructors)
\item Other Reconstructors (6 Reconstructors)
\end{enumerate}
Even within these groups, every Seed-Reconstructor uses unique techniques and functions to reconstruct the seed. These are based on the reverse-engineered source code of the DGA. However, the group names outline the underlying basic principles the reconstruction techniques are based on.\newline

\textbf{Permutators}: \\
The \textit{Permutators} reconstruct the seed of domain names which are permutations of one input seed. Example: The family \textit{VolatileCedar} generates permutations of the domain `getadobeflashplayer.net' such as `egtadobeflashplayer.net'. For these, any domain can be used as seed because the order of the used letters do not matter since all permutations are generated.\\

\textbf{Iterators}:\\
The \textit{Iterators} are applied to domain names that use a base domain string where specific integers or strings are prepended or appended iteratively. Example: The family \textit{BeeBone} generates domains by using the base seed-string "ns1.dnsfor" and appends integers incrementally. For domains of these families the seed-string can be easily separated from the prepended or appended part and extracted.\\

\noindent \textbf{Bruteforcers}:\\
The \textit{Bruteforcers} iterate through all the seeds or combinations of seeds to generate all possible domain names until a set of seeds is found which generates the desired domain names. \\

\noindent \textbf{Smart Bruteforcers}:\\
The \textit{Smart Bruteforcers} iterate over sets of input seeds just like the \textit{Bruteforcers}. However, the \textit{Smart Bruteforcers} exploit features of the DGAs to limit the possible seed space in a way such that bruteforcing can be accomplished faster. For example, the \textit{MakLoader} DGA only uses 31,373 different seeds due to a modulo operation. For this family, the Bruteforcer has to generate domains only for these 31,373 seeds to find the desired domains.\\

\noindent \textbf{Other Reconstructors}:\\
Lastly, the \textit{Other Reconstructors} describe the DGAs for which none of the above groups apply. These Reconstructors take advantage of other features these DGAs exhibit. For example, it may be possible to reconstruct the seed according to the positions of characters within the domain names. Another possibility is to reverse the entire DGA, which is however often not possible due to irreversible actions like bit-shifts or bit-comparisons. \\

The above categorization is not required for certain DGA families because those DGAs do not make use of (deterministic) seeds. Determinism describes the observability and availability of parameters such as the seed. A non-deterministic DGA might, for example, generate domains based on the ticks or milliseconds since system start, which is unpredictable.

There are some DGAs where creating a Seed-Reconstructor is difficult due to usage of multiple irreversible operations, such as bitshifts. Bruteforcing is also no solution due to long runtime or usage of multiple seeds which both increase the reconstruction time to unfeasible durations.

\iffalse
In case of a wrong classification a Seed-Reconstructor will be applied to a set of domains from a different DGA-family (e.g. actual family: "Ramnit", classification: "Tinba", "Tinba"-Reconstructor being applied to "Ramnit"-domains). This will typically lead to an unsuccessful extraction of a seed.
\fi

The Seed-Reconstructors are mainly intended for new campaigns of known DGAs. However, they may also be applicable to new versions (slightly modified source code) depending on the changes made to the DGA.

%-------------------------------------------------------------------------------
\section{Results} \label{sect:results}
%-------------------------------------------------------------------------------
In this section, we give details on the results of the \textit{SESAME} system. We begin by detailing the performance of the underlying ML model, described in Sect. \ref{sect:mlc}. Subsequently, we focus on the two different modules of our \textit{SESAME} system explained in Sect. \ref{sect:batch_classifier} (\textit{Batch Classifier}) and Sect. \ref{sect:seed_extr} (\textit{Seed-Reconstructor}). We analysed DNS-logs for 232 days with \textit{SESAME} and guide the reader through our results. These results are all leading towards the final goal of identifying new DGAs, versions or campaigns and extracting seeds of the latter two.

\begin{figure}[t!]
\includegraphics[width=1.0\columnwidth, trim={145mm 28mm 370mm 19mm},clip]{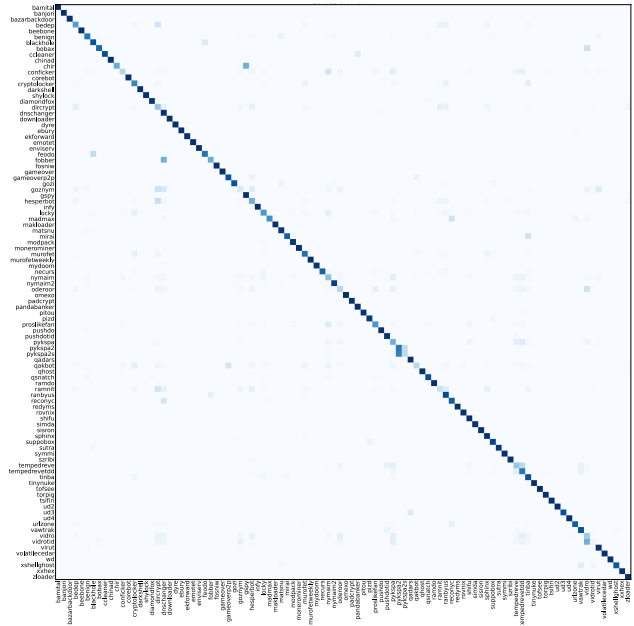}
\caption{Confusion Matrix for the ResNet model. The x-axis shows the predicted families and the y-axis shows the actual families. A perfect model would lead to an identity matrix, coloring the diagonal black and leaving the other fields white.} 
\label{fig:conf_matrix_small}
\end{figure}

\subsection{Machine Learning} \label{results:ML}
First, we evaluated the performance of our machine learning model which is the main tool of the first module, the \textit{Batch Classifier}.
In total, 414,326 unique domains were used in the training of the model. We performed a 5-fold cross validation and determined the Overall Accuracy, (weighted) F1-score, (weighted) Precision and (weighted) Recall. To have a value for comparison, we also give the metrics achieved by the approach of Drichel et al. \cite{2020AachenImbalance} for their ResNet. However, note that they used a different data set, preprocessing, and metrics calculation. While the authors averaged the singular metrics for each DGA-family, we used weighted equations, given in Appendix \ref{apdx: calcs}. We found the following values (the values in parenthesis are taken from \cite{2020AachenImbalance}): F1-score: 0.8409 (0.7868), Precision: 0.8451 (0.7984), Recall: 0.8583 (0.7974), Accuracy: 0.8378 ( - ).

The final model was trained according to Sect. \ref{sect:mlc} and reached an overall accuracy of 83.59\% with a precision of 0.8448, a recall of 0.8602 and an F1-score of 0.8340. The confusion matrix is given in Figure \ref{fig:conf_matrix_small}. An ideal model would show an identity matrix. However, our model is not ideal, thus some deviation can be identified. In total, 97\% of the families had the majority of their domains correctly predicted. This was important, as we used a majority vote to determine the family of a group of domains.

Note that classification correctness may differ between different versions and/or campaigns of one DGA family, i.e. some campaigns may be easier to classify correctly than others. For example, the DGA family \textit{Tinba} has 144 different known sets of seeds. Some of these campaigns purely use the `.com' Top-level domain (TLD) while others use different singular TLDs or multiple TLDs. According to the confusion matrix, our model performs well in predicting the \textit{Tinba} family in general (True Positive Rate of 0.8768). But for the DNS-logs, 32.26\% of the cases in which DGArchive identified a \textit{Tinba} sample, our classifier predicted the family \textit{Ramnit}. 95.37\% of these wrongly classified samples used only the `.com' TLD and 4.45\% of samples used the `.com' and `.net' TLDs. The `.com' TLD is also primarily used by the \textit{Ramnit} family. 

Thus, the metrics derived on the test data set may not represent the same metrics when applied on the DNS-logs. Therefore, we investigate the actual accuracy in Section \ref{result:accuracy}.

\subsection{\textit{SESAME}: Coverage}\label{sect:coverage}
Before explaining the detailed results of the \textit{Batch Classifier}, we give a general overview about the coverage of the analysed data.

\begin{figure}[t!]\centering
\includegraphics[width=\columnwidth]{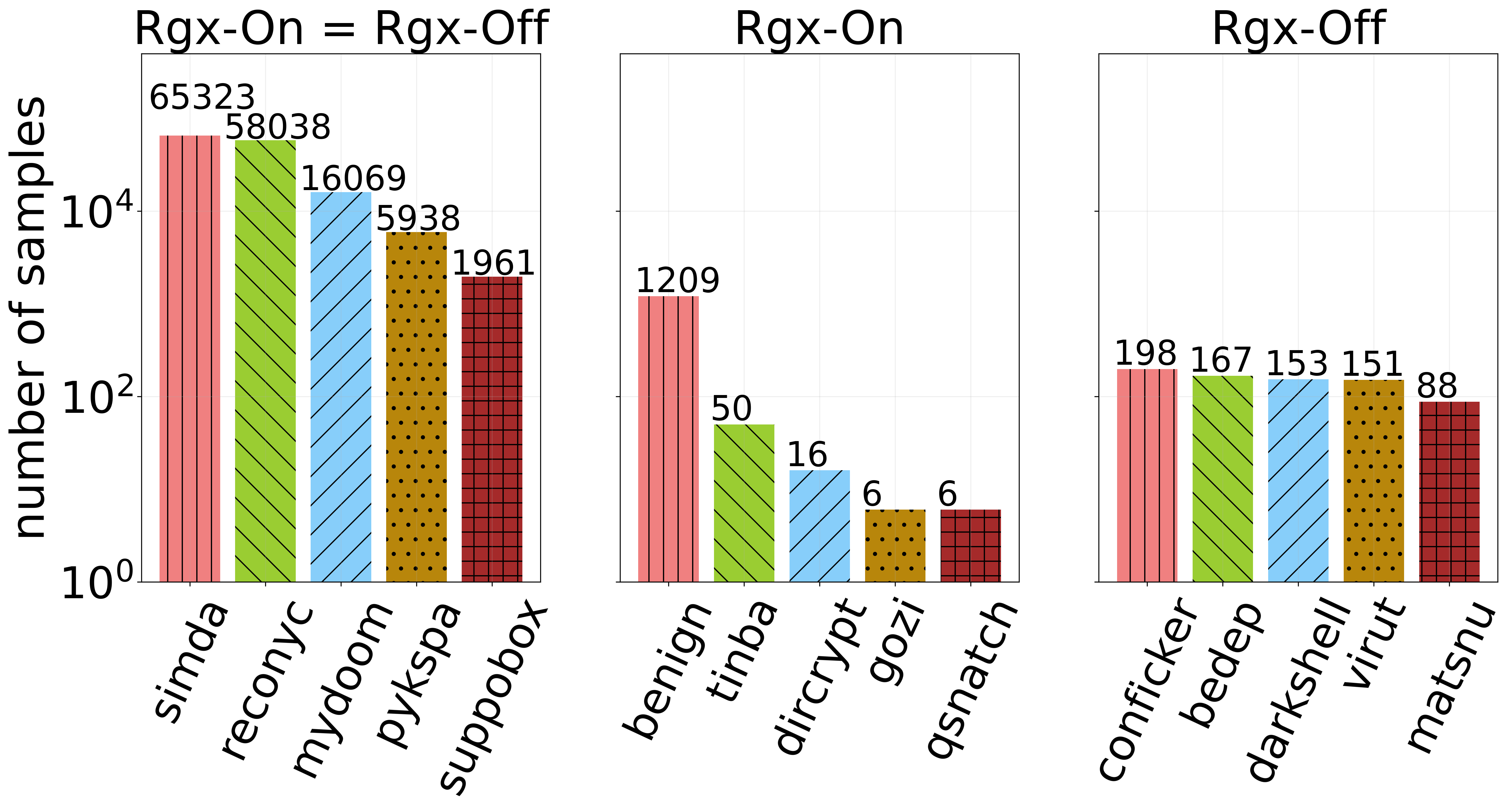}
\caption{Predictions of the \textit{Batch Classifier}. `Rgx-On' and `Rgx-Off' describe whether the \textit{RegEx-Matcher} is utilized. The leftmost graph shows the classifications for which both modes reach the same prediction. Note the logarithmic y-axis.}
\label{fig:results_predictions}
\end{figure}

\textit{SESAME} analysed 232 days of DNS-logs and predicted the DGA-families for 153,472 samples. The results of the classifications depends on the mode the \textit{Batch Classifier} module is operated in. In Sect. \ref{sect:batch_classifier}, we explained how the \textit{RegEx Matcher} feature works and that two predictions are made: one prediction with `matched' RegExes and one prediction without changes. Hence, both predictions can be the same, or they can differ. The five most frequent predicted families for both cases are shown in Figure \ref{fig:results_predictions}. 

\begin{figure}[t!]\centering
\includegraphics[width=0.9\columnwidth]{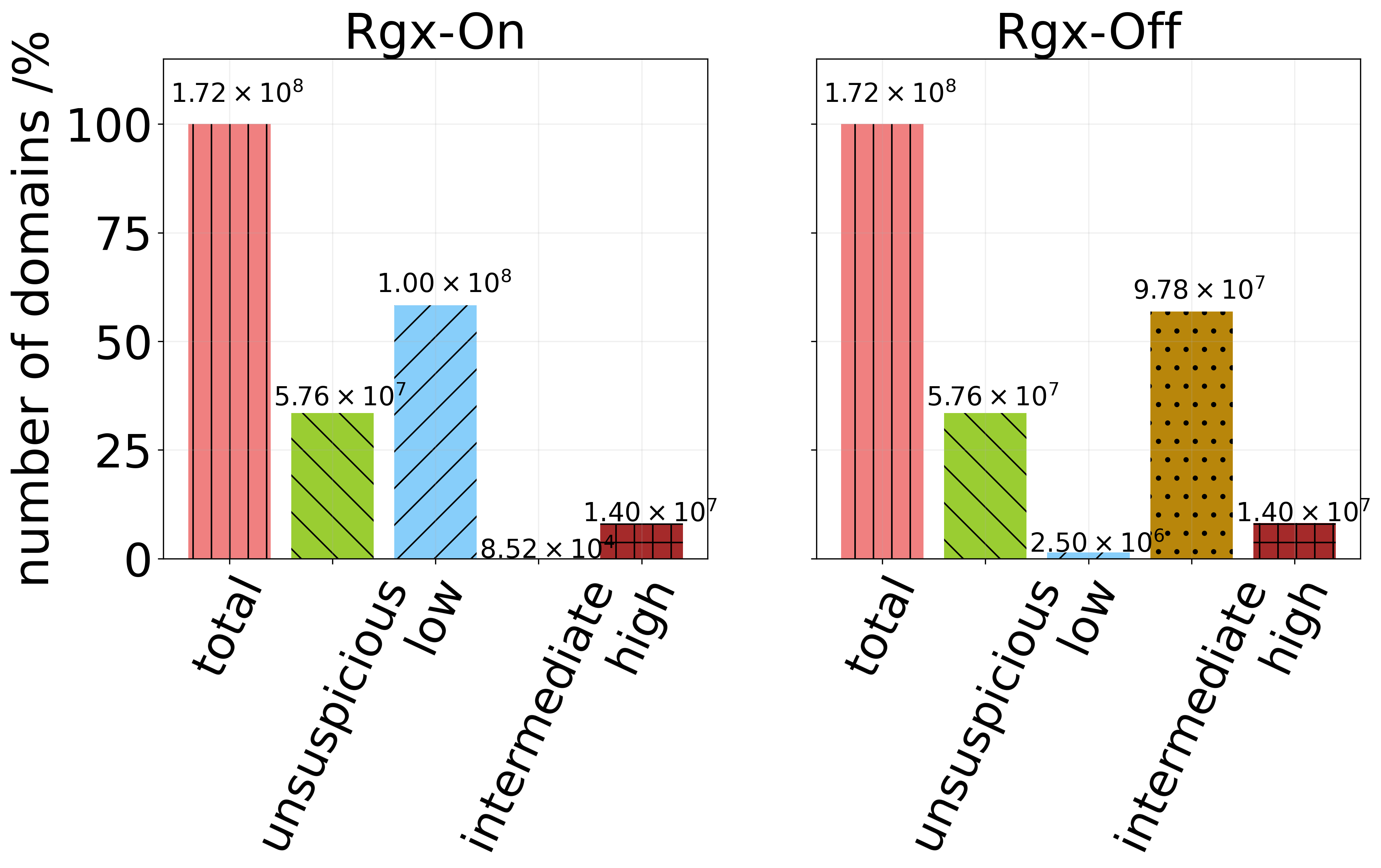}
\caption{Number of classified domains grouped by the calculated suspicion score. `Rgx-On' and `Rgx-Off' describe whether the \textit{RegEx-Matcher} is utilized. Absolute values are given on top of each bar.}
\label{fig:severity}
\end{figure}

The analysed hashes contained over 171 million domains. Figure \ref{fig:severity} summarizes the number of domains according to the suspicion score the respective sample received. We categorized the level of suspicion into 4 classes according to the calculated score:
\begin{itemize}
    \item `unsuspicious': score = 0.0 , 
    \item `slightly suspicious': score $\in$ (0.0 - 5.0] , 
    \item `suspicious': score $\in$ (5.0 - 25.0] , 
    \item `highly suspicious': score $>$ 25.0 .   
\end{itemize}
The lower limit of 5.0 is motivated in Section \ref{sect:ident_tests}. The higher limit of 25.0 is based on empirical test runs.

\subsection{\textit{Batch Classifier}: Accuracy} \label{result:accuracy}
To estimate the performance of our system on DNS-logs, we compared the classifications with the families determined by the database lookup (for which we use DGArchive). Of the 153,472 analysed hashes, 94,233 (61,40\%) contained domains recognised by DGArchive. In this subsection, we evaluate these hashes while the remaining 59,239 samples are further analysed in the subsequent subsections.

\begin{figure}[t!]\centering
\includegraphics[width=0.85\columnwidth]{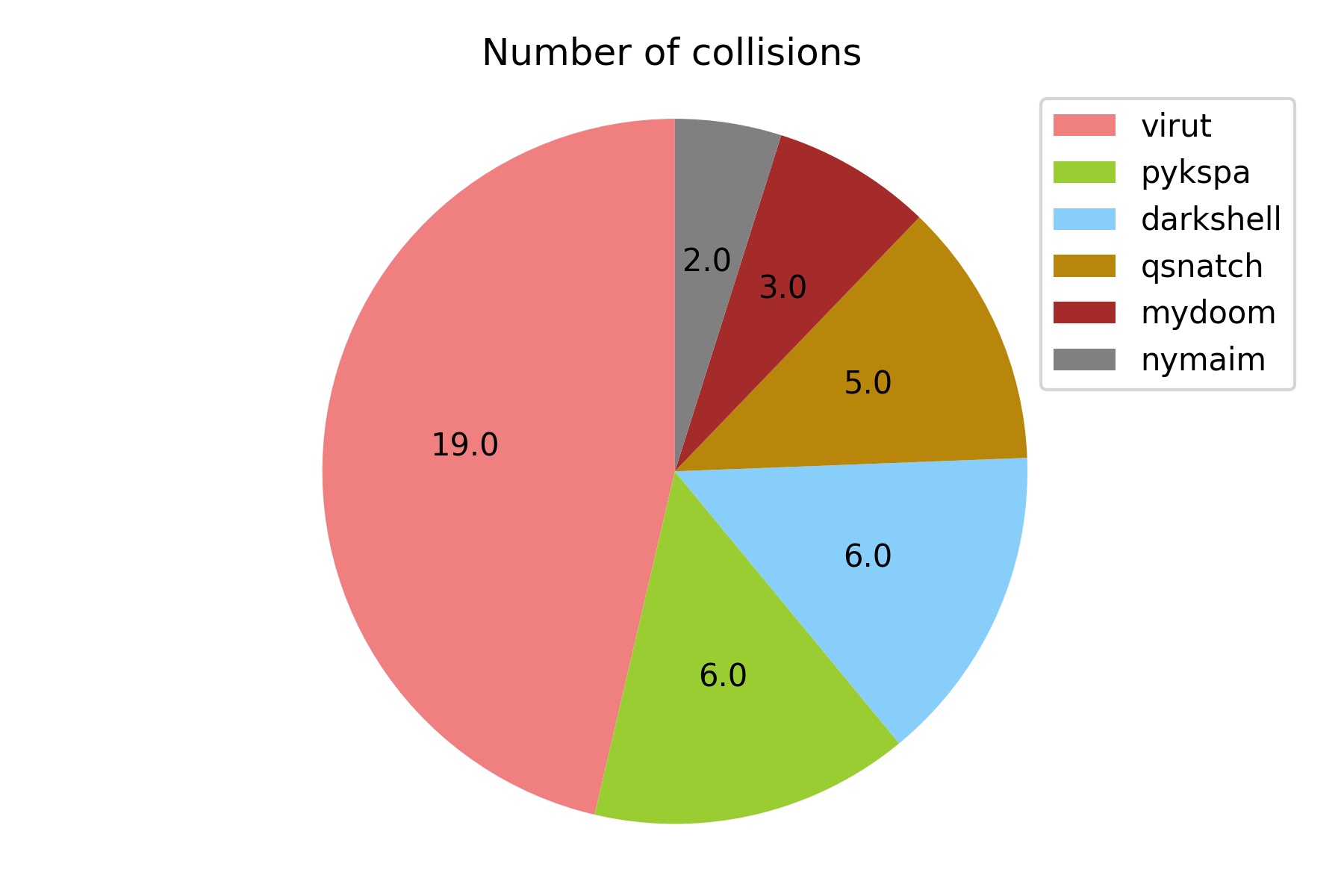}
\caption{Number of samples likely to be collisions. Collisions occur due to overlapping domain name spaces with a resulting wrong classification.}
\label{fig:collisions}
\end{figure}

During the lookup, DGArchive may recognise domains of multiple families within one singular sample. This may happen due to overlapping domain name spaces. These overlaps may also occur for domains of new DGAs or hard-coded domains. Hence, the DGArchive identified families are by no means error-free. However, for the entirety of the analysed hashes, we only identified 41 (0.04\%) of such occurrences. We visualised these collisions in Figure \ref{fig:collisions}. We classified a DGArchive-identification as erroneous or as a collision when only less than either an absolute value of five or 3\% of the analysed domains within a sample were recognised by DGArchive. The majority of collisions happened for the \textit{Virut} family. \textit{Virut} generates short domains of 6 characters, which often overlap with benign domains.

\begin{figure}[t!]\centering{
\includegraphics[width=0.80\columnwidth]{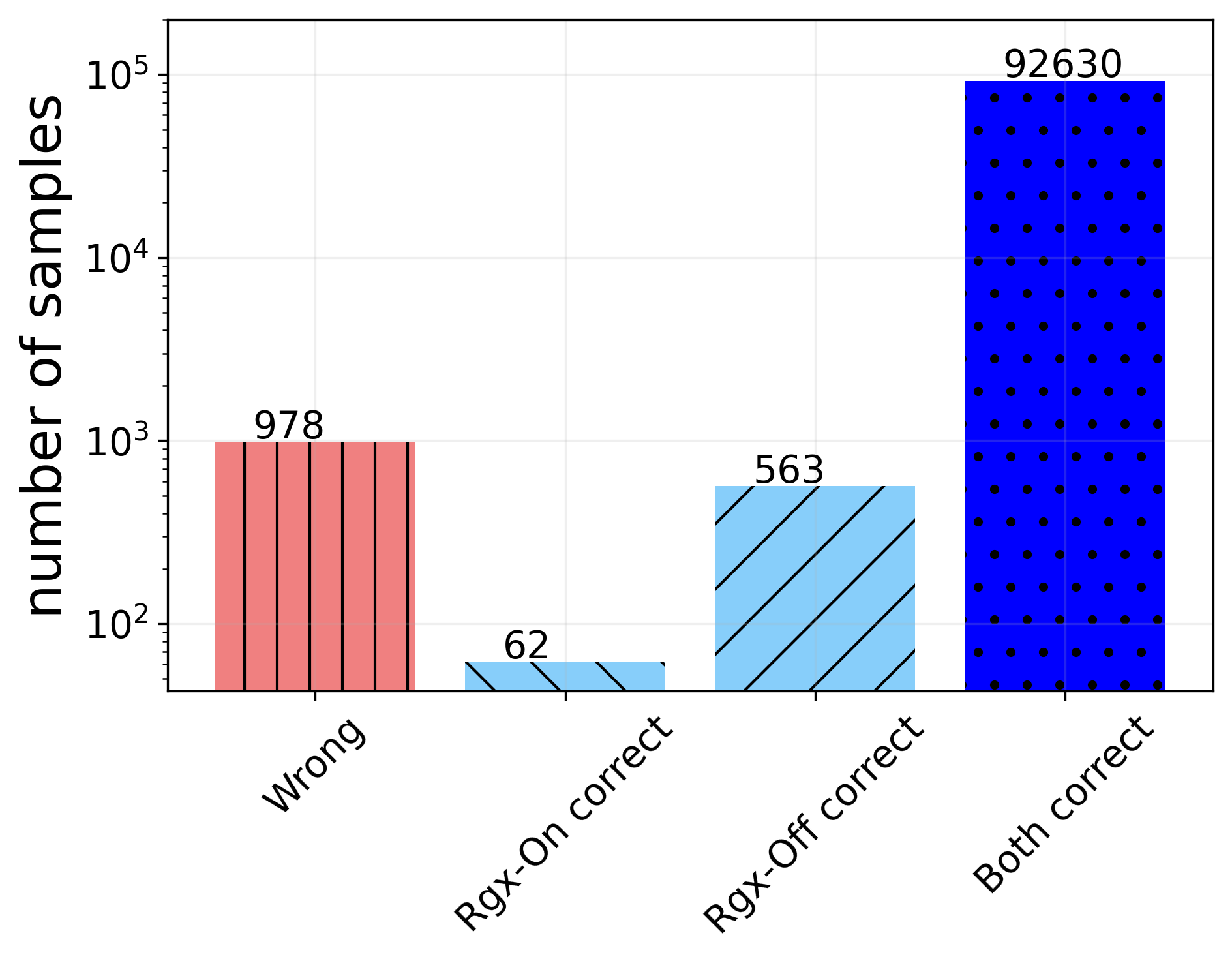}}
\caption{Number of samples identified by DGArchive according to the classifications of \textit{SESAME}. `Wrong' classifications indicate the samples for which both with and without \textit{RegEx-Matcher} no prediction equaled the DGArchive family. The `correct' predictions are shown where one or both of the classifications matched the DGArchive family. Note the logarithmic y-axis.}
\label{fig:results_performance}
\end{figure}

To measure the accuracy, we calculated the number of hashes for which the predictions match the family identified by DGArchive. Note, that since we had two predictions (one with and one without \textit{RegEx-Matcher}, see Sect. \ref{sect:batch_classifier}) we can compare multiple values. We determined how often two correct predictions, only one correct prediction, and no correct predictions were made. The values are visualised in Figure \ref{fig:results_performance}. For 92,630 samples (98.29\%), both predictions matched the family DGArchive identified. For 978 samples (1.04\%), the prediction differed from the DGArchive one. For the remaining 625 samples (0.66\%), only one prediction matched the family identified by DGArchive. This is a good foundation to assume that our model performs reasonably well, at the very least for the campaigns and versions it was trained on. For the cases in which only one prediction was correct, mostly the prediction with the \textit{RegEx-Matcher} disabled is correct. Further analysis showed that with the \textit{RegEx-Matcher} enabled, these particular malware samples were classified as \textit{benign} - the fallback class. This indicates that these samples tried to resolve domains which do not match the actual families' RegEx (e.g. by querying benign domains).

\iffalse
\subsection{\textit{Batch Classifier}: Accuracy over Time}
\textcolor{gray}{It is important to know how often a model has to be re-trained for it to produce satisfying results. Hence, we checked the model's accuracy over time by employing the above-mentioned method of comparing the families identified by DGArchive with our models predictions. To do this, we arbitrarily chose 12 weeks and determined the total number of recognised samples and the number of recognised samples correctly classified by DGArchive. From this, we inferred the accuracy by calculating the ratio of these two values. We visualised the ratios in Figure \ref{fig:results_acc_over_time}. No clear trend can be observed, which indicates that the achieved accuracy does not diminish over time. Therefore, the model does not require frequent re-training.}
%However, we suggest a re-training of the model once 'enough' (which could be just one) new versions, campaigns and/or DGAs become known and should be able to be recognised by the model. 
\fi

\iffalse
\begin{figure}[t!]\centering
\includegraphics[width=0.85\columnwidth]{images/acc_over_time2.png}
\caption{Accuracy of the \textit{Batch Classifier} over time. The number of correct classifications for each week is shown (percentile, values shown are rounded to 2 decimals) along with the starting dates of the randomly evaluated weeks.} % Note that there was no data for the day 2020-10-29.
\label{fig:results_acc_over_time}
\end{figure}
\fi

\subsection{\textit{Batch Classifier}: Identification Tests}\label{sect:ident_tests}
Now that we have established that the classifications are trustworthy, we investigate if it can also serve its intended purpose to separate known from new samples. For this, we applied \textit{SESAME} to a data set which serves as ground truth.
In specific, we generated domains both for seeds which are known and unknown. The unknown seeds are seeds that can potentially be used by an adversary. They might even already be in use without our knowledge.

We generated and tested different numbers of new campaigns for each DGA for which we have a working Seed Reconstructor. In these tests, we created new campaigns which have different numbers of seeds changed from known values. As a practical example, suppose multiple campaigns of DGA \textit{X} are known to us. The DGA uses seeds `A' (an arbitrary integer), `B' (one out of two values) and `C' (one out of two TLDs). In our tests, we generated domains for all possible permutations of seeds.  Thus, one campaign would have only seed `A' changed to an unknown value. Another campaign has seed `C' (the TLDs) changed, and yet another has seed `A' and `C' changed, etc.

In total, we classified domains for 619 sets of seeds. 427 of these were fake seeds and 192 were real seeds as found in databases such as DGArchive. 28 of those 427 fake seeds actually used real seeds and just had the number of generated domains increased. Their suspicion scores depended on how many domains were recognised by DGArchive. Samples with less than a third of the domains recognised ended up with a score above 5.00, which is the limit for a sample to be classified as `suspicious' (see Sect. \ref{sect:coverage}). 19 out of 22 samples that received suspicion scores of less than 5.00 only had increased numbers of domains. The remaining three samples generated enough DGArchive-known domains to receive a small suspicion score. Lastly, all except one real set of seeds was assigned a score higher than 5.00. This happened because the domains generated by this seed were missing in the database dump we used for the lookup. The numbers are also summarized in Table \ref{table:diff_test} for convenience.

This test shows that \textit{SESAME} performs well on differentiating new campaigns from known campaigns. However, when TLDs are changed either with or without using new seeds, for most DGAs, this led to miss-classifications, indicating the importance of the TLD to the model.

\begin{table}[ht]
\centering
\begin{tabular}{ |c|c|c| } 
\hline
	& 	score $>5.00$ 	&  score $\leq5.00$		\\
\hline
total real seeds 	& 1 	& 191	\\
\hline
total fake seeds 	& 405 	& 22	\\
\hline
real, but more domains   & 9 & 19 \\
\hline
\end{tabular}
\caption{Suspicion Scores for the differentiation tests. Samples with fake seeds are mostly labeled `suspicious' i.e. reaches scores of larger than 5.00, while samples with real seeds are mostly labeled `unsuspicious'.}
\label{table:diff_test}
\end{table}

%\iffalse
%\fi

\subsection{\textit{Batch Classifier}: Unknown Domains} \label{sect: unkown}
We showed that \textit{SESAME}'s classification accuracy is good on known samples and we showed that it can identify sets of domains generated by new seeds on a data set that served as ground truth. Next, we analysed the unknown domains from the DNS-logs. Here, we defined `unknown' domains as domains not recognised by DGArchive. We filtered out all malware samples containing domains recognised by DGArchive and ended up with 59,239 samples consisting of 98,032,659 domains.

We found that 58,039 of these samples (97.97\%) were classified as the \textit{Reconyc} family, which generates unpredictable domains. Therefore, no \textit{Reconyc} domains are calculated in DGArchive. Because of the non-deterministic nature of this family, it is difficult to confirm the correctness of these classifications. Reverse engineering every single sample is not feasible because of the large number of \textit{Reconyc} classifications. However, there is an indicative method to decide whether we can trust these results or not. 

We investigated the \textit{Reconyc} predictions from the test set within the confusion matrix (Figure \ref{fig:conf_matrix_small}). The model was able to correctly classify 82.40\% of the \textit{Reconyc} domains. Furthermore, we inspected the extracted RegExes for the DNS-logs samples classified as \textit{Reconyc} and we found that 99.95\% match the exact RegEx of the \textit{Reconyc} family. Accordingly, we believe these \textit{Reconyc} classifications are genuine and trustworthy.

\begin{figure}[t!]
\includegraphics[width=1.0\columnwidth]{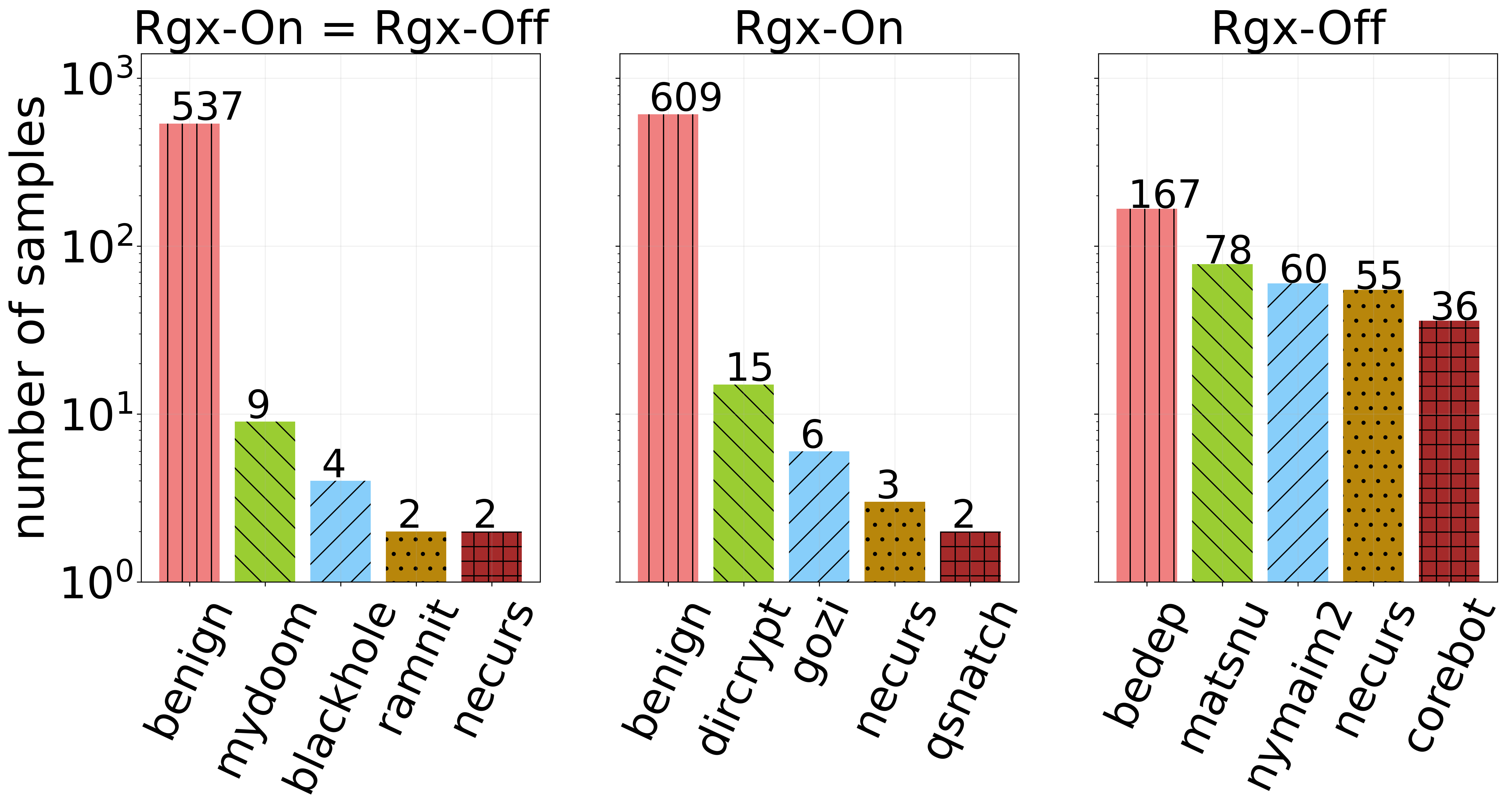}
\caption{Predictions of the \textit{Batch Classifier} for samples unknown to DGArchive. `Rgx-On' and `Rgx-Off' describe whether the \textit{RegEx-Matcher} is utilized. The leftmost graph shows the classifications for which both modes reach the same prediction. We show only the 5 most frequently classified families.}
\label{fig:results_unknown_sample_preds}
\end{figure}

Subsequently, we removed all \textit{Reconyc} classifications from these samples and remained with 1,200 unknown samples (314,304 domains). In Figure \ref{fig:results_unknown_sample_preds} we show the five most frequently classified families for these samples according to the mode of operation of the \textit{RegEx-Matcher}.

For these samples, the classifications of the different modes of the \textit{RegEx-Matcher} were the same for only 46.75\% of the cases (compared to 99.29\% for the samples known by DGArchive). When the \textit{RegEx-Matcher} was utilized, most samples were classified as `benign' (the fallback class) whereas the classifications were way more evenly distributed between different DGA families without the \textit{RegEx-Matcher}. This could indicate that many of these samples do not follow the scheme of known DGAs. Since the input is assumed to be generated by malware, this indicates that these samples are likely to contain mostly new DGAs and sets of hard-coded domains. However, they may possibly also contain new versions and even new campaigns.

Nevertheless, unknown DGAs, versions and/or campaigns can also be found within the set of data containing domains recognised by DGArchive. This may occur due to overlapping domain names or small numbers of unique generated domains of an algorithm.

%\iffalse
%\fi

\subsection{\textit{Seed Reconstructor}: Coverage}
Before we apply the Seed Reconstructors we give a general overview about the \textit{Seed Reconstructor} module.

At the time of writing, we have Seed Reconstructors for 30 different DGA families. Another 13 DGA families do not require seed reconstruction, as no seeds are utilized. These families are the following: \textit{Bamital}, \textit{Bedep}\footnote{Seeds are weekly currency exchange courses}, \textit{Blackhole}, \textit{CCleaner}, \textit{Cryptolocker}, \textit{DNSBenchmark}, \textit{DNSChanger}, \textit{Dyre}, \textit{Ekforward}, \textit{GameoverP2P}, \textit{Gspy}, \textit{Modpack} and \textit{Pizd}. This covers 43.43\% of the DGAs. 

16.16\% of the DGA families are difficult cases for which we still need to come up with a reconstruction scheme. These DGAs mostly use hashing algorithms while using multiple seeds or expensive calculations which harden them against bruteforcing. 

Thus, 43 of the DGAs are already covered, for 16 the work is temporarily halted and 40 are still under investigation.

\subsection{\textit{Seed Reconstructor}: Identifications}
With the \textit{Batch Classifier}, we were able to assign scores to the sets of domains. Using this score we can now identify `suspicious' samples for which we have functional Seed Reconstructors.

Overall, there are 3,142 `suspicious' samples for which we have implemented Seed Reconstructors. 3,022 of these were samples that contained domains recognised by DGArchive as part of the families \textit{MyDoom} (2839, 93.9\%), \textit{Darkshell} (162, 5.4\%)), \textit{Pushdo}, \textit{Nymaim}, \textit{Suppobox} and \textit{Conficker}. As indicated, most of these samples belong to the families \textit{MyDoom} and \textit{Darkshell}. The remaining 120 samples contained no domains recognised by DGArchive and were classified as ten different DGA families: \textit{Blackhole}, \textit{Chinad}, \textit{Dircrypt}, \textit{Locky}, \textit{MyDoom}, \textit{Necurs}, \textit{Pushdo}, \textit{Qadars}, \textit{Qakbot} and \textit{Suppobox}.

While many of the DGArchive unknown samples were wrong classifications and contained hard-coded domains, we were able to extract a new \textit{Dircrypt} seed, as well as a \textit{MyDoom} seed. The reason the \textit{MyDoom} samples were tagged as suspicious is, because the DGArchive implementation generated too few domains. Furthermore, we discovered new {Darkshell} and {Suppobox} seeds.

In Section \ref{sect: unkown} we reduced the number of unknown samples to 1200. 1199 are classified as suspicious, of which 120 have functional Seed Reconstructors. The remaining 1079 samples, as well as the samples mentioned here, for which we were not able to extract seeds, we investigated further in the next subsection.

\subsection{\textit{SESAME}: Unknown DGAs, Versions and Campaigns}\label{subsect: RE}
In the previous sections, we were able to reduce the number of `suspicious' samples. By further grouping the samples that have the same RegExes, we can further reduce the cardinality.

Finally, we end up with over 64 'suspicious' samples that may potentially contain a DGA. While 3 samples are yet to be reverse engineered, we were able to identify 32 samples using hard-coded domains. For 10 samples we were not yet able to determine whether the malware sample uses hard-coded domains or a time independent DGA. Furthermore, we identified 2 different samples which we believe to contain different DGAs. However, this hypothesis still requires more investigation. Finally, the remaining 17 groups of domains turned out to be generated by DGAs.

Of these, we have identified 4 new DGAs. 1 of these turned out to be the \textit{Shylock} DGA (2 Versions). We attributed 2 of those DGAs to the families of \textit{Emudbot}/\textit{Skybot} and \textit{GrabBot}. The last one is undocumented and thus entirely new. We called it \textit{Hiasm} (based on what we assume to be the malware authors PC name). 5 DGAs were new versions of known DGAs (\textit{Hesperbot}, \textit{Suppobox}, \textit{Darkshell} and 2x \textit{Tsifiri}). We were able to extract 25 new seeds for those \textit{Darkshell} versions. 2 DGAs were new campaigns of the DGAs of \textit{Dircrypt} and \textit{Pykspa}. 1 DGA was the known \textit{Mydoom} DGA which was incorrectly implemented in the DGA database and thus generated too few domains. Additionally, we found 2 DGAs of the families \textit{Oderoor} and \textit{Pushdo}, where we are still investigating whether they are new version or campaigns. We also analysed a \textit{Recoync} DGA which scored a particularly high score compared to other \textit{Recoync}-classified samples. However, it appears to contain the known DGA. Lastly, another \textit{Dircrypt} version was identified because this version was not covered by the database dump we used, but by the time we identified this version, the seed was already implemented in the database.

%For many of these, possible multiple versions and/or campaigns were identified. However, reverse-engineering is required to confirm or reject the hypothesis of these cases being DGAs. The number of 64 potentially new DGAs is very large compared to the number of 99 currently known DGA families. 

%-------------------------------------------------------------------------------
\section{Discussion} \label{sect:discussion}
%-------------------------------------------------------------------------------
In this section, we discuss our results as well as their implications. We also give statements about the current state of the work and how it can be further improved.

\subsection{Model Accuracy}\label{discuss:model_accuracy}
We implemented the ResNet model proposed by Drichel et al. \cite{2020AachenLstm} and achieved an overall accuracy of 83.89\%. Under the assumption that the samples recognised by DGArchive can be used as labeled AGDs, we found that 98.29\% of classifications of the DNS-logs are correct (see Sect. \ref{result:accuracy}).

\subsection{Seed Reconstruction and its limitations}
Every Seed-Reconstructor is handcrafted for one DGA-family. This requires an initial investment of effort once and then future seeds can be reconstructed effortless. Our approach is advantageous to extractors, as instead of analysing a malware sample, it only requires sets of domains. We showed the proof of concept for 30 different DGA families. However, there are a few general limitations.

Firstly, there can be extended run times, especially when iterating over many seeds. To keep computation times short, for some DGAs, the default seed space to be iterated over is kept in ranges which were seen on currently known versions of those DGAs. This may lead to some seeds not being properly reconstructed.

Secondly, many DGAs use the system time as one seed. We assume the system time to correspond to the date of malware execution of the DNS-logs. 

Thirdly, some of the analysed malware contains domains not generated by a DGA. Hence, including all domains of a sample for the Seed Reconstructor may not be beneficial. We limited the included domains for the Seed Reconstruction to a smaller number of domains which is adjustable for every DGA. This value depends on the DGA and is defaulted to values between 5 and 100.

Some of these limitations may lead to an inconclusive execution of an otherwise functional Seed Reconstructor. There were not enough cases in the DNS data to estimate how much influence these limitations carry in terms of successfully reconstructing seeds. Furthermore, for some DGAs we could not identify a weakness that could be exploited. In many of these cases, a hashing function and/or many different seeds are used which makes the Seed reconstruction very difficult or time-consuming.

\subsection{Confirming potentially new DGAs, Versions and Campaigns}
One of the main goals of \textit{SESAME} is to reconstruct seeds of currently unknown versions and campaigns of already known DGAs. However, we also find groups of domains potentially belonging to completely unknown DGA families. We found 64 groups of domains possibly generated by new campaigns, versions, new DGAs or which might be hard-coded in the malware. To confirm how each sample's domains were generated, these samples were reverse engineered.

We found 32 malware samples using hard-coded domains and 17 samples using DGAs. Of these, 4 were entirely new to us at the time of discovery. The remaining ones were new versions or campaigns from known DGAs. Additionally, 10 samples which use either hard-coded domains or time independent DGAs, and 5 samples, of which we believe 2 are using DGAs, require further reverse engineering.

This already shows that our system works well to identify `suspicious' samples which are likely to contain new DGAs, versions or campaigns. %It shows how the work of a reverse engineer can be notably reduced as instead of 153,472 malware samples only 64 had to be reverse engineered.

\subsection{Robustness}
It takes a one-time effort to create a \textit{Seed Reconstructor} which is able to extract seeds from domains for a certain DGA (and all of its known versions). Changing the seed more regularly, will not be of much use as our identification tests (Sec. \ref{sect:ident_tests}) showed that new campaigns and versions can be detected, and the new seeds can simply be extracted.

Should an adversary purposely alter the algorithm itself in order to generate new domains and try to evade our \textit{SESAME} system, then the corresponding \textit{Seed Reconstructor} will most likely also require updates. In this case, every one-time effort of an adversary leads to a one-time effort of a security specialist. One can argue that it is less effort for the adversary than for the security specialist to implement updates. However, our system then would have already increased the workload of an adversary. That said, our system is designed to reduce the effort of the security specialists by reducing the amount of samples that need manual inspection. Having identified new DGA versions also helps the security researchers to further shorten the required effort. Experienced researchers are able to extract seeds from malware faster when they know which DGA is implemented.

%-------------------------------------------------------------------------------
\section{Conclusion} \label{sect:conclusion}
%-------------------------------------------------------------------------------

We created a system named \textit{SESAME} which is based on two modules. The \textit{Batch Classifier} module determines the particular DGA family of groups of domains along with a score measuring the level of novelty. For the determination of the DGA family, we trained a multi-class classification model based on a residual neural net. A high score indicates a large difference between the schemes of the analysed domains and the ones known to the model. In the second step, the \textit{Seed Reconstructor} module automatically reconstructs seeds from DGA samples which were assigned scores larger than a threshold score. These seeds can then in turn be used to calculate past and future domains.

With our approach, we were able to distinguish unknown DGAs, versions and campaigns, as well as, sets of hard-coded domains from known DGAs.

%We propose a novel approach to reconstruct seeds from domain names in order save time and effort typically required for reverse engineering. 

%Due to the uniqueness of each DGA, we create a unique Seed Reconstructor for every DGA. While this takes a one-time effort and works for some DGAs better than for others, we give a working proof of concept of this method.

After analysing 232 days of DNS-logs from malware sandbox runs, we identified 64 suspicious malware samples, possibly utilizing entirely new DGAs. Some of these contain multiple versions and/or multiple campaigns. While 15 samples require further reverse engineering, we were already able to discover 17 DGAs and 32 malware samples with hard-coded domains. Thus, we were able to break down 153,472 analysed malware samples to 64, which had to be reverse engineered, to confirm that our system works as intended.

We showed a working proof of concept for our \textit{SESAME} system. While there is still room for improvement, we showed that the combination of our \textit{Batch Classifier} and our \textit{Seed Reconstruction} module are able to successfully reconstruct seeds without manual effort (during the analysis and based solely on domain names. To the best of our knowledge, there are no other works which focus on recovering seeds from domain names.

Therefore, we fulfilled our goal of combining the strengths of ML and database approaches by providing a system with state-of-the-art ML results that is also able to automatically reconstruct seeds with which future domains can be pre-calculated.

\iffalse

%-------------------------------------------------------------------------------
\section*{Acknowledgments}
%-------------------------------------------------------------------------------

Thanks to Daniel Plohmann for solid advice, questions, answers and comments as well as granting access to DGArchive.\\
Thanks to Shadowserver for doing the work they do, as well as generously providing us access to the Sandbox Logs.\cite{shadowserver}\\
\fi

%-------------------------------------------------------------------------------

%\bibliographystyle{ACM-Reference-Format}
\bibliographystyle{plain}
\bibliography{SESAME}{}

%-------------------------------------------------------------------------------
\appendix
%-------------------------------------------------------------------------------

\section{Sandbox Logs Data Details}\label{apdx:log_details}
Here, we give a more detailed overview about the statistics of the sandbox logs. This analysis is useful to understand how to best achieve the goal of detecting unknown campaigns of known DGAs.\\
Every 24 hours a new log file is created. The number of unique malware executed per day varies. Shadowserver aggregates their samples from multiple sources and depending on these, more or fewer samples are executed per day. We analyse the feeds for 28 arbitrarily chosen days and show the results to get an overview about the feeds. This is done in an attempt to not crowd the plots unnecessarily while still being representative for the entirety of the data.\\ %Analyzing the entirety of the feeds would have crowded the plots and ruined any possibility of getting an overview about the statistics.

%-------------------------------------------------------------------------------
\subsubsection{Number of Domains}
%-------------------------------------------------------------------------------

\begin{figure}[t!]
\includegraphics[width=\columnwidth]{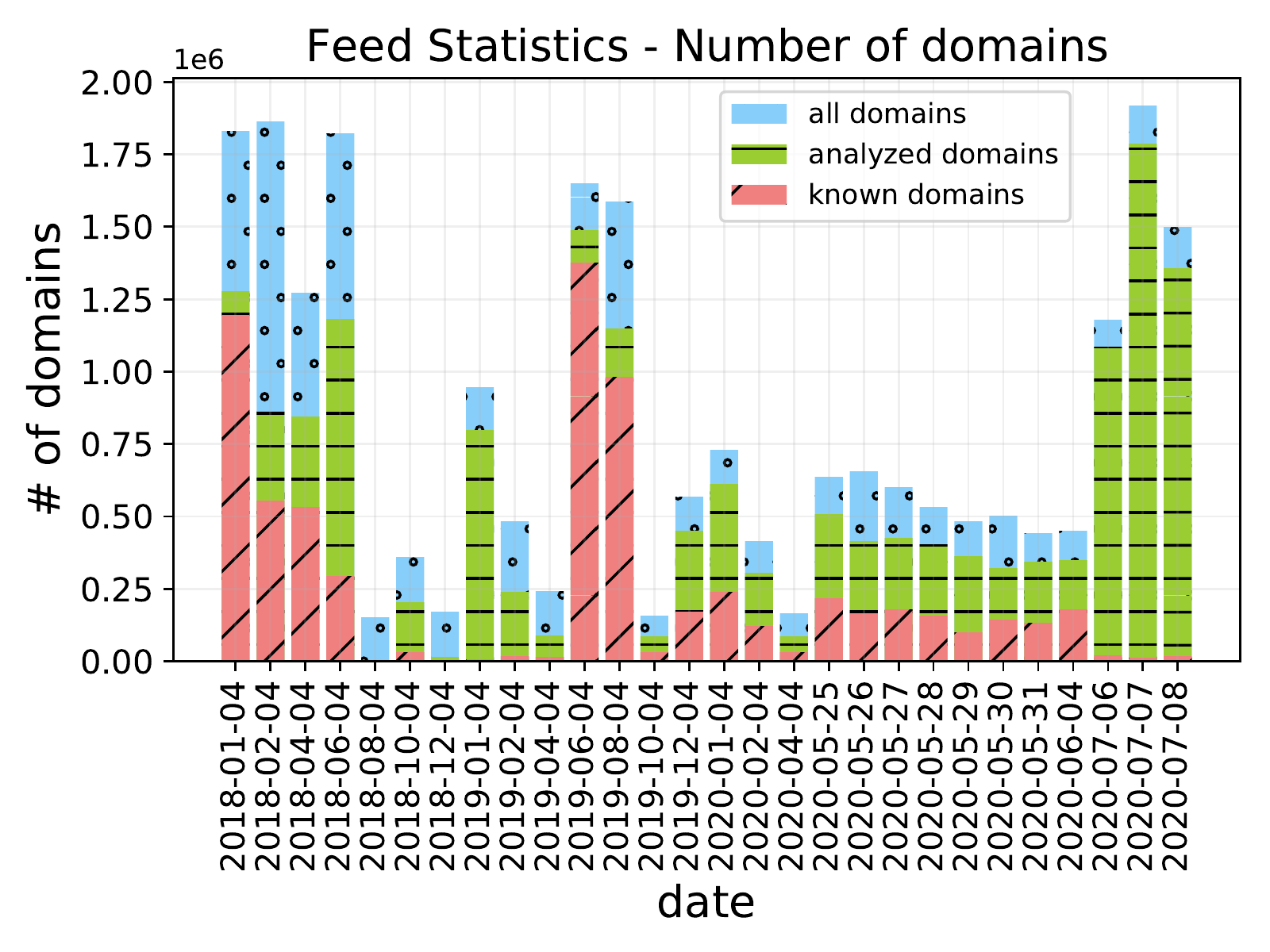}
\caption{Visualisation of the total number of domain names (blue), only analysed domain names (green) and the ones recognised by DGArchive (red) per day.}
\label{fig:feed_domains}
\end{figure}

Different malware generates different amounts of domains per time interval. Some malware first tries to resolve one or multiple benign domains (e.g. `bing.com') to assess the network connectivity. In Figure \ref{fig:feed_domains} we show the amounts of domains (blue) per log file. A few filters are applied (see section \ref{sect:batch_classifier}) to the logs to remove e.g. broken domain names. Thus, the number of domains which are actually used for the analysis is reduced (green). Lastly, the number of domains which are already known by DGArchive are highlighted (red). All these three values vary over time. Though, for consecutive days the change seems to be less drastic (see the week from 2020-05-25 to 2020-05-31). For more recent logs, we see a large amount of domains while the amount of domains recognised by DGArchive is extremely low. Note that the DGArchive domains are already pre-calculated for these dates, therefore this is not an effect of missing domains but an actual feature. Using the \textit{Batch Classifier} (see Sect. \ref{sect:batch_classifier}), we identify 95.6\% of the investigated hashes (for date 2020-07-07) as the \textit{Reconyc} family whose DGA is unpredictable and therefore not stored in DGArchive.\\
In total, the feeds used for the statistical overview contained 23.296.012 Domains of which 17.031.635 (73.11\%) remain after filtering and of which 6.904.586 are known by DGArchive (29.64\%). \\
This clearly shows that even after applying our current knowledge about most known DGAs (assuming DGArchive contains most of the currently known DGAs), there still remain many unknown domains which requires another tool than just a blocklist. Manual inspection is infeasible for these amounts of domains, and so is reverse engineering every single sample. Hence, our usage of ML to break down the number of samples to the actually interesting ones. Subsequently, these can be further examined for seed reconstruction.\\

%-------------------------------------------------------------------------------
\subsubsection{Feed Responses and Feed Types} \label{apdx:feed_types}
%-------------------------------------------------------------------------------

\begin{figure}[t!]
\includegraphics[width=\columnwidth]{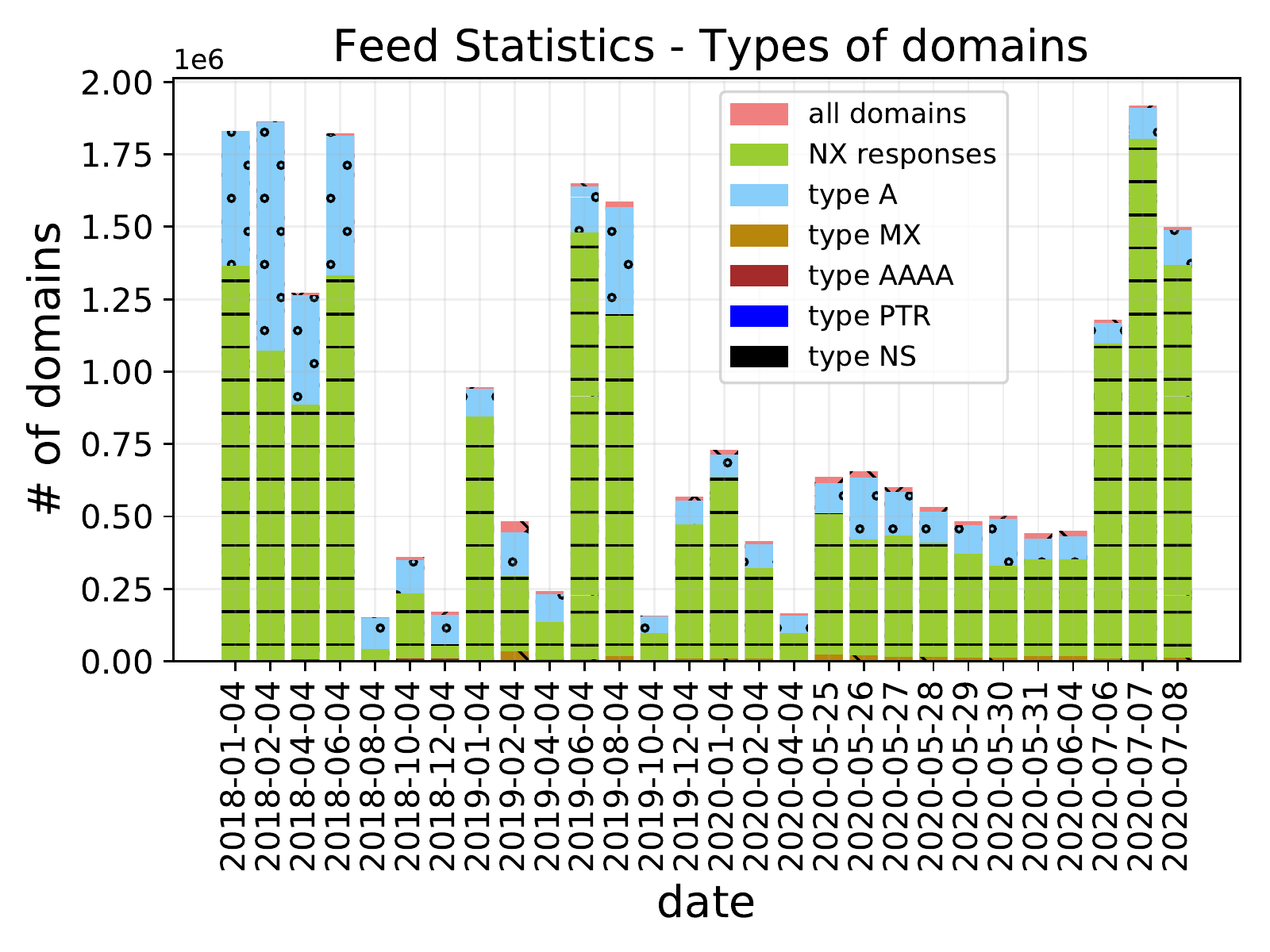}
\caption{The number of domains per DNS-type per day are shown, as well as the number of NXDomain responses for any of these types.}
\label{fig:feed_responses}
\end{figure}

Every domain lookup comes with a certain DNS-type and a certain response. In our data, all NXDomain responses are returned as `0.0.0.0'-responses and as such these two responses are used interchangeably. Under the assumption that most of the DGA-generated domains are not actually being registered the majority of responses should receive the NXDomain response. Figure \ref{fig:feed_responses} shows that this assumption holds true. The Figure illustrates the fraction of NXDomain responses for each day as well as the amount of domains belonging to the different DNS-types. Most of the DNS-types are the `A'-type which is also to be expected of DGA-generated domains.\\

%-------------------------------------------------------------------------------
\subsubsection{Number of Unique Domains}
%-------------------------------------------------------------------------------

\begin{figure}[t!]
\includegraphics[width=\columnwidth]{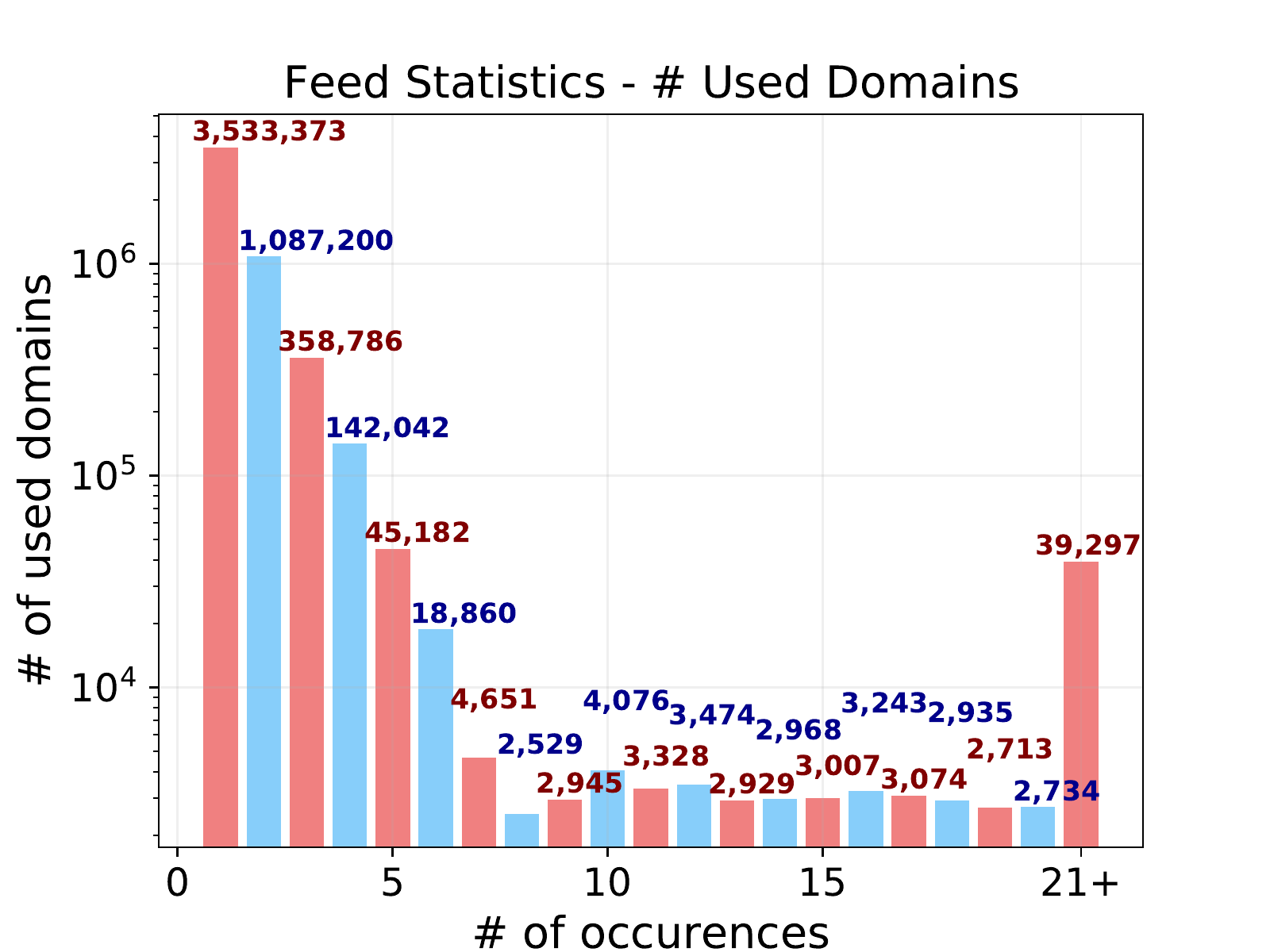}
\caption{The number of used domains along with their occurrences are shown. All domains that occur 21 or more times are grouped up in the last bar for better visibility. Note the logarithmic y-axis.}
\label{fig:feed_used_domains}
\end{figure}

Every malware sample generates domains according to the underlying DGA family. Therefore, some DGAs generate (partly) the same domains even when executed on different days. Also, different malware samples which use the same DGA will in most cases generate the same domains when executed on the same day. Malware often tries to test their network connection by querying a benign domain before querying the actual DGA domains. For this check, many malware samples use the same benign domains. The most used benign domain in the analysed logs is `bing.com' which is queried 9921 times throughout the feeds investigated here. Hence, we expect to see many domains occurring more than once. In Figure \ref{fig:feed_used_domains} we can observe exactly what we expected. Note the sudden increase in domains that appear `21+' times. The reason for this is, that all domains appearing 21 or more times are grouped up in this bar. It is possible that domains only re-occur because the same hash is executed multiple times, therefore we investigate the re-appearance of hashes next.

%-------------------------------------------------------------------------------
\subsubsection{Number of Unique Hashes}
%-------------------------------------------------------------------------------

\begin{figure}[t!]
\includegraphics[width=\columnwidth]{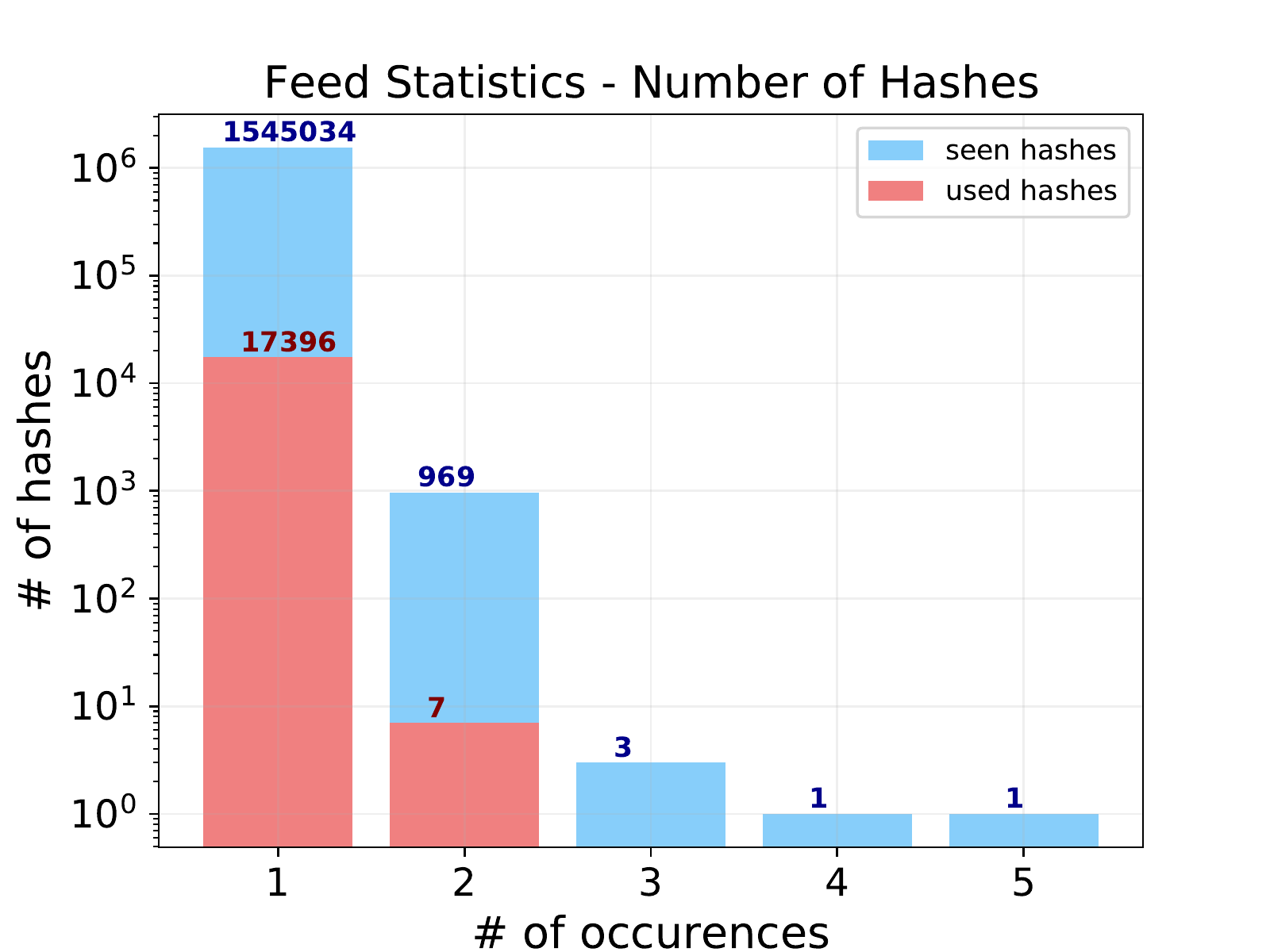}
\caption{This visualisation illustrates the number of hashes and how often they (re-)appear in the analysed logs. Note the logarithmic y-axis.}
\label{fig:feed_hashes}
\end{figure}

As previously mentioned, different numbers of samples are executed every day. In Figure \ref{fig:feed_hashes} we visualise the amount of hashes and their occurrences for all the analysed logs (shown in green in Figure \ref{fig:feed_domains}). Overall, out of 1.546.008 unique hashes only 17.403 remain after applying filters for invalid domains (mentioned in Sect. \ref{sect:batch_classifier_preprocesing}). 969 hashes appear twice throughout the analysed logs. Very few hashes appear more than twice. Though, only seven of these remain after applying our filters.\\
Clearly most (99.96\%) of the hashes which pass the filter criteria, only occur once. This proves that re-occurring domain names are not a result of analysing the same sample multiple times.

%-------------------------------------------------------------------------------
\subsubsection{Sample Execution Durations}
%-------------------------------------------------------------------------------

\begin{figure}[t!]
\includegraphics[width=\columnwidth]{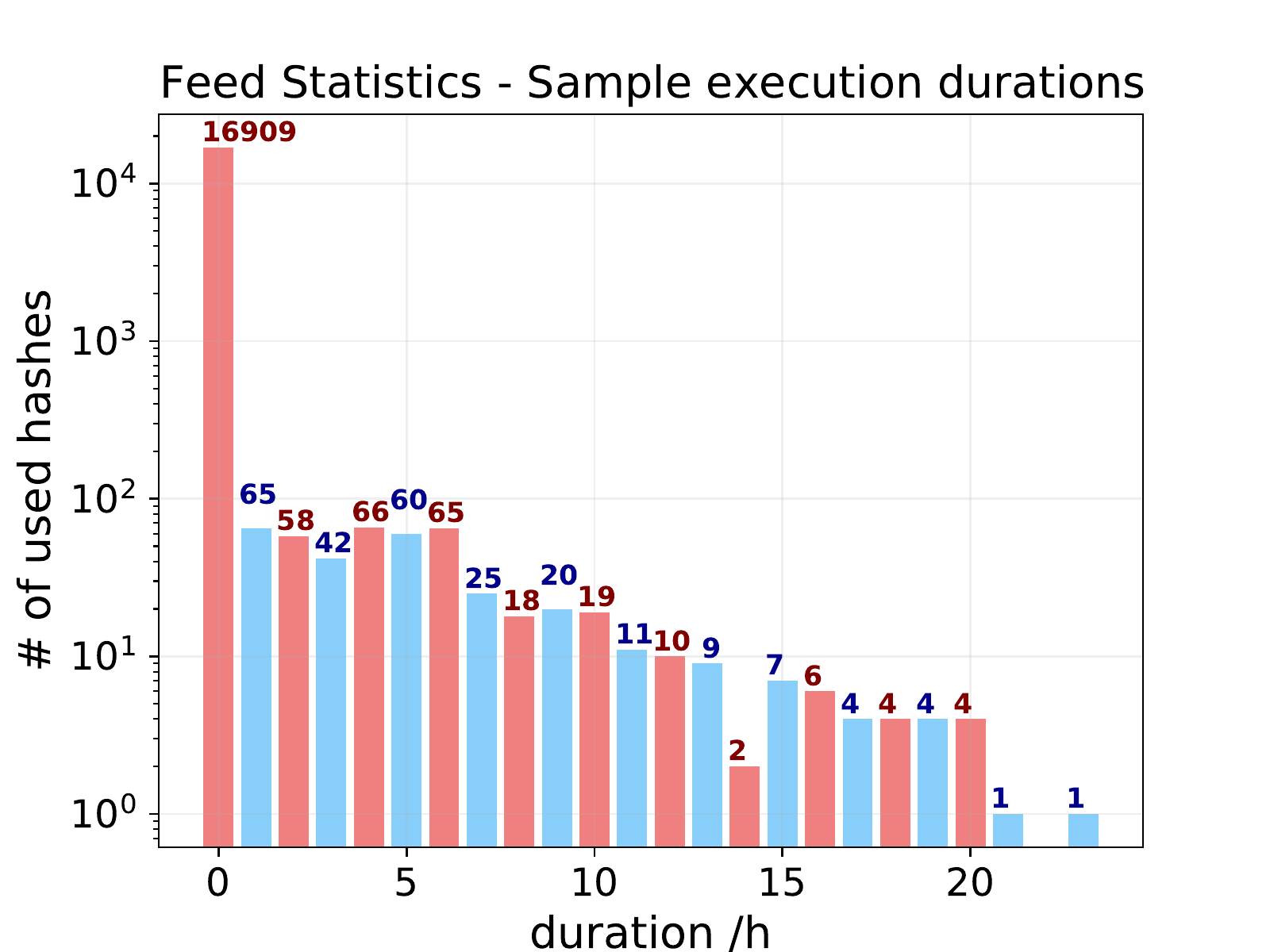}
\caption{The amount of samples grouped according to their execution duration. Note the logarithmic y-axis.}
\label{fig:feed_exec_durations}
\end{figure}

The different malware samples are executed for different durations. In Figure \ref{fig:feed_exec_durations} we show the amount of analysed hashes grouped by their duration times in hours. Note the logarithmic y-axis. `Duration' is to be understood as the time from the first the last occurrence of a certain md5 hash. A duration of zero implies a time between zero seconds and 1 hour at most. Most of the hashes are executed for a short duration in the order of minutes, which explains the large first spike. However, some samples are executed over a longer period of time. The longest duration amounts to 23 hours and 31 minutes. However, investigations showed for this particular sample that about 800 domains were queried around 00:14 AM to 00:16 AM and another 2000 domains were queried between 11:44 PM and 11:46 PM. \\
According to ShadowServer the automatic malware execution is time-boxed to a maximum of 10 minutes. However, if a certain malware triggers a yara/AV/snort signature, then this sample will be rescheduled for another run outside of the sandbox. This could explain the observed behaviour of long `execution' durations.\\
Other observed behaviours for samples with long durations showed that a single `benign' domain may be looked up early in the morning but the DGA domains are only queried later in the night. This may indicate a usage of some kind of Sandbox detection for these particular samples. The opposite has also been observed: In the early hours many domains are queried and one final `benign' domain is resolved late at night.\\
The actual durations during which DGA-domains are looked up, tend to stay within the range of a few tens of seconds to minutes.

%-------------------------------------------------------------------------------
\subsection{Calculations}\label{apdx: calcs}
%-------------------------------------------------------------------------------

Here we give the calculations for the metrics of the Machine Learning model:

\iffalse
\noindent \begin{math}
    ACC_{\text{Overall}}=\frac{\sum_{i=1}^{|C|}TP_i}{\sum_{i=1}^{|C|}TP_{i} + FN_{i}}  \\ 
    \text{Precision}=\sum_{i=1}^{|C|}\frac{TP_i}{TP_i+FP_i} \cdot \frac{N_i}{N} \\
    \text{Recall}=\sum_{i=1}^{|C|}\frac{TP_i}{TP_i+FN_i} \cdot \frac{N_i}{N}\\ 
    F_{1}=\sum_{i=1}^{|C|}\frac{2TP_i}{2TP_i+FP_i+FN_i}\cdot \frac{N_i}{N}\\
    \text{TP: True Positives, TN: True Negatives} \\
    \text{FP: False Positives, FN: False Negatives} \\
\end{math}
\fi

\begin{flalign}
    \vspace{0.4cm}
    & ACC_{\text{Overall}} =\frac{\sum_{i=1}^{|C|}TP_i}{\sum_{i=1}^{|C|}TP_{i} + FN_{i}} \label{eq:acc} \\ 
    \vspace{0.4cm}
    & \text{Precision} =\sum_{i=1}^{|C|}\frac{TP_i}{TP_i+FP_i} \cdot \frac{N_i}{N} \\
    \vspace{0.4cm}
    & \text{Recall} =\sum_{i=1}^{|C|}\frac{TP_i}{TP_i+FN_i} \cdot \frac{N_i}{N}\\ 
    \vspace{0.2cm}
    & F_{1} =\sum_{i=1}^{|C|}\frac{2TP_i}{2TP_i+FP_i+FN_i}\cdot \frac{N_i}{N} \label{eq:F1}
\end{flalign}

\noindent TP: True Positives, TN: True Negatives \\
FP: False Positives, FN: False Negatives \\

\noindent With the number of classes $C$, the number of domains per class $N_i = FN_i + TP_i$ and the total number of domains \\$N$ = $\sum_{i=1}^{|C|}(FN_i + TP_i)$. 
The FN, FP, TN and TP are all calculated per class. To understand how they are determined, see the following example: 
\begin{itemize}
\item True Positive: classifying a \textit{Conficker} domain as \textit{Conficker}.
\item True Negative: classifying a domain from any other class not as \textit{Conficker}.
\item False Negative: classifying a \textit{Conficker} domain as any other class.
\item False Positive: classifying a domain from any other class as \textit{Conficker}.
\end{itemize}

%-------------------------------------------------------------------------------
\subsection{Technical Challenges}\label{apdx: challenges}
%-------------------------------------------------------------------------------

\begin{table}[H]
\centering

\begin{tabular}{|M{3.5cm}|M{3.5cm}|}
    \hline
    \textbf{technical challenge} & \textbf{solution (reference)} \\
    \hline
    select best dataset to train model & select 5k domains per family acc., to certain rules (\ref{subsect:mldata})  \\
    \hline
    find benign dataset with no malicious domains & use Tranco data + select only first 500k domains (\ref{subsub_benign_data})  \\
    \hline
    evaluate the system on a real dataset & apply system to ShadowServer data (\ref{subsect:sandbox})\\
    \hline
    evaluate performance of the ML model & perform 5-fold CV (\ref{sect:model_eval}) \\
    \hline
    best way to identify already known DGAs within analysed groups of domains & compare with database (\ref{sect:batch_classifier_dgarchive_crosscheck})\\
    \hline
    assign a score to judge novelty of a group of domains & multiple calculations (\ref{sect:batch_classifier_sus_score}) \\
    \hline
    best way to assign correct family to a group of domains & majority vote (\ref{sect:batch_classifier_vote}) + 2 approaches during classification based on DGA-RegEx (\ref{sect:batch_classifier_classification}) \\
    \hline
    evaluate performance of \textit{SESAME} on new campaigns & create domains with fake seeds and measure detection rate (\ref{sect:ident_tests}) \\
    \hline
    best way to see differences between groups of domains assigned the same family & RegEx extractor for easy RegEx comparison (\ref{sect:batch_classifier_rgx_extract}) \\
    \hline
    RegEx extractor & parse group of domain names and extract the used alphabet and lengths \\
    \hline
    a way to recover seeds from the available information & Seed Reconstructors (\ref{sect:seed_extr})\\
    \hline
    confirm \textit{SESAME}'s functionality & Manual Reverse Engineering (\ref{subsect: RE})\\
    \hline
\end{tabular}%
\caption{List of all technical challenges and how they were solved (with references to more detailed explanations).}
\label{table:tech_chals}
\end{table}

\end{document}